\documentclass[12pt, draftclsnofoot,onecolumn]{IEEEtran}

\usepackage{graphicx,subcaption,xcolor,setspace,amsthm,cite,booktabs,epstopdf,dblfloatfix,blindtext,amsthm}
\usepackage{url}
\usepackage{algorithm,algcompatible,upgreek,acro}

\usepackage[binary-units,detect-weight=true, detect-family=true]{siunitx}
\DeclareSIUnit \bitspersecond {bps}
\newcommand{\centrallocation}{/Users/AnumAli/Dropbox/STUDY/Central} 

\usepackage{amsmath,amssymb}

\newcommand{\bbC}{{\mathbb{C}}}

\newcommand{\bbE}{{\mathbb{E}}}

\newcommand{\bbR}{{\mathbb{R}}}

\newcommand{\ba}{{\mathbf{a}}}

\newcommand{\bd}{{\mathbf{d}}}

\newcommand{\bff}{{\mathbf{f}}}

\newcommand{\br}{{\mathbf{r}}}

\newcommand{\bw}{{\mathbf{w}}}
\newcommand{\bx}{{\mathbf{x}}}
\newcommand{\by}{{\mathbf{y}}}

\newcommand{\bzero}{{\mathbf{0}}}

\newcommand{\bF}{{\mathbf{F}}}

\newcommand{\bH}{{\mathbf{H}}}
\newcommand{\bI}{{\mathbf{I}}}

\newcommand{\bN}{{\mathbf{N}}}

\newcommand{\bR}{{\mathbf{R}}}

\newcommand{\bT}{{\mathbf{T}}}
\newcommand{\bU}{{\mathbf{U}}}

\newcommand{\bW}{{\mathbf{W}}}
\newcommand{\bX}{{\mathbf{X}}}
\newcommand{\bY}{{\mathbf{Y}}}

\newcommand{\rmc}{{\mathrm{c}}}

\newcommand{\rme}{{\mathrm{e}}}

\newcommand{\rmm}{{\mathrm{m}}}

\newcommand{\rmp}{{\mathrm{p}}}

\newcommand{\rmr}{{\mathrm{r}}}
\newcommand{\rms}{{\mathrm{s}}}

\newcommand{\rmB}{{\mathrm{B}}}

\newcommand{\rmF}{{\mathrm{F}}}

\newcommand{\rmR}{{\mathrm{R}}}
\newcommand{\rmS}{{\mathrm{S}}}

\newcommand{\rmU}{{\mathrm{U}}}
\newcommand{\rmV}{{\mathrm{V}}}

\newcommand{\cC}{\mathcal{C}}

\newcommand{\cI}{\mathcal{I}}

\newcommand{\cL}{\mathcal{L}}

\newcommand{\cN}{\mathcal{N}}

\newcommand{\cT}{\mathcal{T}}
\newcommand{\cU}{\mathcal{U}}

\newcommand{\bsfn}{\boldsymbol{\mathsf{n}}}

\newcommand{\bsfs}{\boldsymbol{\mathsf{s}}}

\newcommand{\bsfx}{\boldsymbol{\mathsf{x}}}
\newcommand{\bsfy}{\boldsymbol{\mathsf{y}}}

\newcommand{\bsfF}{\boldsymbol{\mathsf{F}}}

\newcommand{\bsfH}{\boldsymbol{\mathsf{H}}}

\newcommand{\bsfR}{\boldsymbol{\mathsf{R}}}

\newcommand{\bsfW}{\boldsymbol{\mathsf{W}}}

\newcommand{\transp}{{\sf T}}
\newcommand{\compj}{{\rm j}}
\newcommand{\tr}{{\rm tr}}

\newcommand{\diag}{{\rm diag}}

\makeatletter
\def\munderbar#1{\underline{\sbox\tw@{$#1$}\dp\tw@\z@\box\tw@}}
\makeatother
\newcommand{\dBm}{\SI{}{\decibel}\rmm}

\DeclareAcronym{3GPP}{
  short=3GPP,
  long=3rd generation partnership project
}
\DeclareAcronym{ADC}{
  short=ADC,
  long=analog-to-digital converter
}
\DeclareAcronym{AMP}{
  short=AMP,
  long=approximate message passing
}
\DeclareAcronym{AoA}{
  short=AoA,
  long=angle-of-arrival
}
\DeclareAcronym{AoD}{
  short=AoD,
  long=angle-of-departure
}
\DeclareAcronym{APS}{
  short=APS,
  long=azimuth power spectrum
}
\DeclareAcronym{AV}{
  short=AV,
  long=autonomous vehicle
}
\DeclareAcronym{BS}{
  short=BS,
  long=base station
}
\DeclareAcronym{BSM}{
  short=BSM,
  long=basic safety message
}
\DeclareAcronym{CP}{
  short=CP,
  long=cyclic-prefix
}
\DeclareAcronym{DFT}{
  short=DFT,
  long=discrete Fourier transform
}
\DeclareAcronym{DL}{
  short=DL,
  long=downlink
}
\DeclareAcronym{DSRC}{
  short=DSRC,
  long=dedicated short-range communication
}
\DeclareAcronym{FDD}{
  short=FDD,
  long=frequency division duplex
}
\DeclareAcronym{FMCW}{
  short=FMCW,
  long=frequency modulated continuous wave
}
\DeclareAcronym{FoV}{
  short=FoV,
  long=field-of-view
}
\DeclareAcronym{GNSS}{
  short=GNSS,
  long=global navigation satellite system
}
\DeclareAcronym{LIDAR}{
  short=LIDAR,
  long=Light detection and ranging
}
\DeclareAcronym{LOS}{
  short=LOS,
  long=line-of-sight
}
\DeclareAcronym{LPF}{
  short=LPF,
  long=low pass filter
}
\DeclareAcronym{LTE}{
  short=LTE,
  long=long term evolution
}
\DeclareAcronym{MIMO}{
  short=MIMO,
  long=multiple-input multiple-output
}
\DeclareAcronym{MRR}{
  short=MRR,
  long=medium range radar
}
\DeclareAcronym{NLOS}{
  short=NLOS,
  long=non-line-of-sight
}
\DeclareAcronym{NR}{
  short=NR,
  long=new radio
}
\DeclareAcronym{OFDM}{
  short=OFDM,
  long=orthogonal frequency-division multiplexing
}
\DeclareAcronym{ppm}{
  short=ppm,
  long=parts-per-million
}
\DeclareAcronym{RMS}{
  short=RMS,
  long=root-mean-square
}
\DeclareAcronym{RPE}{
  short=RPE,
  long=relative precoding efficiency
}
\DeclareAcronym{RSU}{
  short=RSU,
  long=roadside unit
}
\DeclareAcronym{SNR}{
  short=SNR,
  long=signal-to-noise ratio
}
\DeclareAcronym{UL}{
  short=UL,
  long=uplink
}

\DeclareAcronym{ULA}{
  short=ULA,
  long=uniform linear array
}
\DeclareAcronym{V2I}{
  short=V2I,
  long=vehicle-to-infrastructure
}
\DeclareAcronym{V2V}{
  short=V2V,
  long=vehicle-to-vehicle
}
\DeclareAcronym{V2X}{
  short=V2X,
  long=vehicle-to-everything
}
\DeclareAcronym{VRU}{
  short=VRU,
  long=vulnerable road user
}

\newcommand{\Ns}{N_{\rms}}

\newcommand{\NRSU}{N_{\rmR\rmS\rmU}}
\newcommand{\NV}{N_{\rmV}}
\newcommand{\MRSU}{M_{\rmR\rmS\rmU}}
\newcommand{\MV}{M_{\rmV}}
\newcommand{\bFBB}{\bF_{\rmB\rmB}}
\newcommand{\bFRF}{\bF_{\rmR\rmF}}
\newcommand{\bWBB}{\bW_{\rmB\rmB}}
\newcommand{\bWRF}{\bW_{\rmR\rmF}}
\newcommand{\alpharc}{\alpha_{r_c}}
\newcommand{\baRSU}{\ba_{\rmR\rmS\rmU}}
\newcommand{\baV}{\ba_{\rmV}}

\newcommand{\bsfRRSU}{\bsfR_{\rmR\rmS\rmU}}
\newcommand{\hbsfRRSU}{\hat{\bsfR}_{\rmR\rmS\rmU}}

\newcommand{\fr}{f_\rmr}

\newcommand{\Tr}{T_{\rmr}}
\newcommand{\Tc}{T_{\rmc}}

\newcommand{\Delf}{\Delta f}
\begin{document}
\bstctlcite{IEEEmax3beforeetal}
\title{Passive Radar at the Roadside Unit to Configure Millimeter Wave Vehicle-to-Infrastructure Links}
\author{Anum Ali, {\it Member, IEEE}, Nuria Gonz\'alez-Prelcic, {\it Senior Member, IEEE}, and, Amitava Ghosh, {\it Fellow, IEEE}
\thanks{This work was supported by a gift from Nokia.}
\thanks{A. Ali is with the Department of Electrical and Computer Engineering, The University of Texas at Austin, Austin, TX 78712-1687, USA \mbox{(e-mail: anumali@utexas.edu)}.}
\thanks{N. Gonz\'alez-Prelcic is with the Department of Electrical and Computer Engineering, The University of Texas at Austin, Austin, TX 78712-1687, USA, and also with the Signal Theory and Communications Department, University of Vigo, 36310 Vigo, Spain \mbox{(e-mail: ngprelcic@utexas.edu)}.}
\thanks{A. Ghosh is with Nokia Bell Labs, Naperville, IL 60563-1594 USA (e-mail: amitava.ghosh@nokia-bell-labs.com)}
}

\maketitle

%
\begin{abstract}
Millimeter wave (mmWave) vehicular channels are highly dynamic, and the communication link needs to be reconfigured frequently. In this work, we propose to use a passive radar receiver at the roadside unit to reduce the training overhead of establishing an mmWave communication link. Specifically, the passive radar will tap the transmissions from the automotive radars of the vehicles on the road. The spatial covariance of the received radar signals will be estimated and used to establish the communication link. We propose a simplified radar receiver that does not require the transmitted waveform as a reference. To leverage the radar information for beamforming, the radar azimuth power spectrum (APS) and the communication APS should be similar. We outline a radar covariance correction strategy to increase the similarity between the radar and communication APS. We also propose a metric to compare the similarity of the radar and communication APS that has a connection with the achievable rate. We present simulation results based on ray-tracing data. The results show that: (i) covariance correction improves the similarity of radar and communication APS, and (ii) the radar-assisted strategy significantly reduces the training overhead, being particularly useful in non-line-of-sight scenarios.
\end{abstract}
\section{Introduction}\label{sec:intro}
\IEEEPARstart{H}{igh} data-rate communication is possible at millimeter wave (mmWave) frequencies, owing to the large bandwidth~\cite{Rappaport2013Millimeter,Pi2016Millimeter}. Low pre-beamforming \ac{SNR}, however, poses several challenges in establishing reliable mmWave links. Highly directional communication employing large antenna arrays at both sides of the communication link can overcome the low \ac{SNR}. Large antenna systems, however, will be inefficient, or even infeasible, from a hardware, cost and energy consumption point of view, if a dedicated RF-chain with a high-resolution \ac{ADC} is used with each antenna element. As such, MIMO architectures based on either low-resolution \acp{ADC}~\cite{Mo2015Capacity,Li2017Channel}, or a small number of RF-chains have to be used~\cite{ElAyach2014Spatially,Alkhateeb2013Hybrid,Molisch2017Hybrid}. These architectures, though low cost and energy efficient, make the link configuration more difficult. The primary reason is that the channel is not accessible at the baseband. In low-resolution architectures, the channel at the baseband is quantized. With limited RF-chains, the baseband observes a low-dimensional projection of the channel at the RF front-end. Therefore, link configuration in energy efficient architectures requires a large training overhead. This problem is further compounded in highly mobile scenarios, like \ac{V2X} communication, where channel changes frequently and rapid link re-configuration is necessary.

In this work, we propose to use a passive radar at the \ac{RSU} to configure the mmWave link. The passive radar receiver array at the \ac{RSU} taps the signals transmitted by the automotive radars on the ego-vehicle.  The mmWave communication channels and the radar received signals stem from the same environment. Therefore, there is bound to be similarity in the spatial information embedded in the radar received signal and the spatial characteristics of the communication channels. As such, we exploit the spatial covariance of the radar received signals for mmWave link configuration. 
\subsection{Contributions}
The main contributions of this paper are as follows:
\begin{itemize}
\item We propose to use a passive radar at the \ac{RSU}. The passive radar at the \ac{RSU} will tap the radar signals transmitted by the automotive radars mounted on the ego-vehicle. The spatial covariance of the radar signals received at the \ac{RSU} will be used to configure the mmWave link.
\item We design a simplified radar receiver architecture that does not require the transmitted waveform as a reference. We show that the spatial covariance of the signals in the simplified architecture is the same as the spatial covariance with perfect waveform knowledge. Due to the lack of waveform knowledge, the range and Doppler cannot be recovered using the proposed architecture. The range and Doppler, however, are not relevant when using radar to configure the communications link.
\item We outline a strategy to correct the bias in \acf{FMCW} radar. In~\cite{Kim2018Joint}, it was shown that the angle estimation in \ac{FMCW} radar is biased. We note that a similar bias appears in \ac{FDD} systems, where the \ac{UL} covariance is used to configure the \ac{DL}. After establishing this connection, we propose a strategy initially used for \ac{FDD} covariance correction~\cite{Jordan2009Conversion}, to correct the bias in \ac{FMCW} radars.
\item We propose a similarity metric to quantitatively compare two power spectra. In order to use the radar information for configuring the mmWave links, it is necessary to understand the congruence (or similarity) of the spatial information provided by radar and the spatial characteristic of the mmWave channel. Intuitively, by congruence, we mean the similarity in the \acf{APS} of radar and communication. We show that, in certain cases, the proposed similarity metric is identical to the \ac{RPE}, i.e., a commonly used metric to measure the accuracy of covariance estimation in literature~\cite{Park2019Spatial,Haghighatshoar2017Massive,Park2018Spatial}. Further, in~\cite{Park2019Spatial}, the \ac{RPE} was related to the rate. As such, establishing a connection between the proposed metric and \ac{RPE} also implies a connection between the proposed similarity metric and rate.
\item We develop a ray-tracing setup in Wireless Insite~\cite{WI} to test the proposed radar aided mmWave link-establishment strategies. Our ray-tracing setup is consistent with \acs{3GPP} \ac{V2X} evaluation methodology~\cite{3GPP37885}. 
\end{itemize}
\subsection{Prior work}
Several sources of out-of-band information were considered for mmWave communication systems. In~\cite{Nitsche2015Steering,Hashemi2018Out,Ali2018Millimeter,Ali2018Spatial}, spatial information extracted from sub-6 GHz channels was used to acquire information about the mmWave channel. For vehicular mmWave communications, the location of the vehicle was used to reduce the beam-training overhead~\cite{Garcia2016Location,Va2017Position}. \Ac{LIDAR} data was also used to detect the \ac{LOS} between the \ac{RSU} and the vehicles, and subsequently to reduce the beam-training overhead~\cite{Klautau2019LIDAR}. In~\cite{Brambilla2019Inertial}, information from inertial sensors mounted on the antenna arrays was used for beam-tracking in \ac{V2V} communication systems.  

Prior work on using location~\cite{Garcia2016Location,Va2017Position}, \ac{LIDAR}~\cite{Klautau2019LIDAR}, and inertial sensors~\cite{Brambilla2019Inertial} is limited only to \ac{LOS} links. The sub-6 GHz assisted strategies in~\cite{Nitsche2015Steering,Hashemi2018Out} are also limited to \ac{LOS}. Though the strategies proposed in~\cite{Ali2018Millimeter,Ali2018Spatial} can be used in \acf{NLOS} links, they require the state of the sub-6 GHz and the mmWave link to be the same. That is to say, either both sub-6 GHz and mmWave links are \ac{LOS}, or both are \ac{NLOS}. In practice, a link is considered \ac{LOS}, if more than $80\%$ of the first Fresnel zone is unobstructed. The size of the Fresnel zone depends on the wavelength, and it is possible that the Fresnel zone clarity criterion is satisfied at sub-6 GHz and not at mmWave. This implies that it is possible to have \ac{LOS} sub-6 GHz and \ac{NLOS} mmWave link, i.e., a scenario where the strategies proposed in~\cite{Ali2018Millimeter,Ali2018Spatial} may not work well.
 
There is also some prior work on using radar for mmWave communications. In~\cite{Gonzalez-Prelcic2016Radar}, the radar covariance was used to design the hybrid analog-digital precoders. In~\cite{Chen2018DoA,Ali2019Millimeter}, the location of the vehicle was estimated using radar, and later this information was used to reduce the mmWave training overhead. In~\cite{Muns2017Beam}, a joint communication-radar system based on IEEE 802.11ad physical layer frames was proposed. Once the location of the vehicle was determined through the radar operation, it was used to reduce the training overhead. In~\cite{Simic2016RadMAC}, radar-based vehicle tracking was used to avoid blockage by preemptively switching to a link that is predicted to be unblocked. For an indoor environment, the location of all targets was estimated using radar in~\cite{Jiao2019Millimeter}. This information was then used to predict the channel as the users move.

As with~\cite{Garcia2016Location,Va2017Position}, the location assisted strategies proposed in~\cite{Chen2018DoA,Ali2019Millimeter,Muns2017Beam,Simic2016RadMAC} are limited only to \ac{LOS} links. Similarly, in~\cite{Jiao2019Millimeter}, it is not possible to recover the \ac{NLOS} targets, and hence their contribution to the channel corresponding to a given user. Radar covariance used in~\cite{Gonzalez-Prelcic2016Radar} provides more information than location. In this work, we also use radar covariance to aid mmWave link establishment. As such, among the prior work,~\cite{Gonzalez-Prelcic2016Radar} can be considered the most similar to this work. The key difference between this paper and~\cite{Gonzalez-Prelcic2016Radar} is that in this work we use a passive radar, whereas~\cite{Gonzalez-Prelcic2016Radar} used an active radar. This difference has two implications. First, mounting an active radar on the \ac{RSU} to aid mmWave link establishment implies power cost. This power cost is remarkably reduced in passive radars as no signal is transmitted. Second, the active radar mounted on the \ac{RSU} is also limited to the \ac{LOS} as the \ac{NLOS} ego-vehicle cannot be detected. In contrast, as we tap the signals transmitted by the ego-vehicle, our approach works even in \ac{NLOS} scenarios. This claim is verified in the simulation section using ray-tracing data.

There are also several technical novelties unique to our work. These include: (i) a simplified radar receiver architecture for passive radar, (ii) correcting the angle estimation bias in \ac{FMCW} radars, and (iii) a similarity metric to assess the similarity of radar and communication \ac{APS} which is connected to the achievable rate.

The rest of this paper is organized as follows: In Section~\ref{sec:V2I}, we discuss the general \ac{V2I} communication setup and the communication system model. We discuss the radar system model, the proposed simplified radar receiver, and the strategy to correct the bias in \ac{FMCW} radar in Section~\ref{sec:rad}. We outline the metric to compare the similarity of radar and communication in Section~\ref{sec:congruence}. Next, we provide the simulation results in Section~\ref{sec:simres}, and finally we conclude the paper in Section~\ref{sec:conc}, outlining also the directions for future work.

\textbf{Notation:} We use the following notation throughout the paper. Bold lowercase $\bx$ is used for column vectors, bold uppercase $\bX$ is used for matrices, non-bold letters $x$, $X$ are used for scalars. $[\bx]_i$ and $[\bX]_{i,j}$, denote $i$th entry of $\bx$ and entry at the $i$th row and $j$th column of $\bX$, respectively. We use serif font, e.g., $\bsfx$, for the frequency-domain variables. Superscript $\transp$, $\ast$ and $\dagger$ represent the transpose, conjugate transpose and pseudo inverse, respectively. $\bzero$ and $\bI$ denote the zero vector and identity matrix respectively. $\cC\cN(\bx,\bX)$ denotes a complex circularly symmetric Gaussian random vector with mean $\bx$ and covariance $\bX$, and $\cU[a,b]$ is a Uniform random variable with support $[a,b]$. We use $\bbE[\cdot]$ and $\|\!\cdot\!\|_\rmF$ to denote expectation and Frobenius norm, respectively.

\section{V2I communication setup}\label{sec:V2I}
\begin{figure}[h!]
\centering
\includegraphics[width=0.7\textwidth]{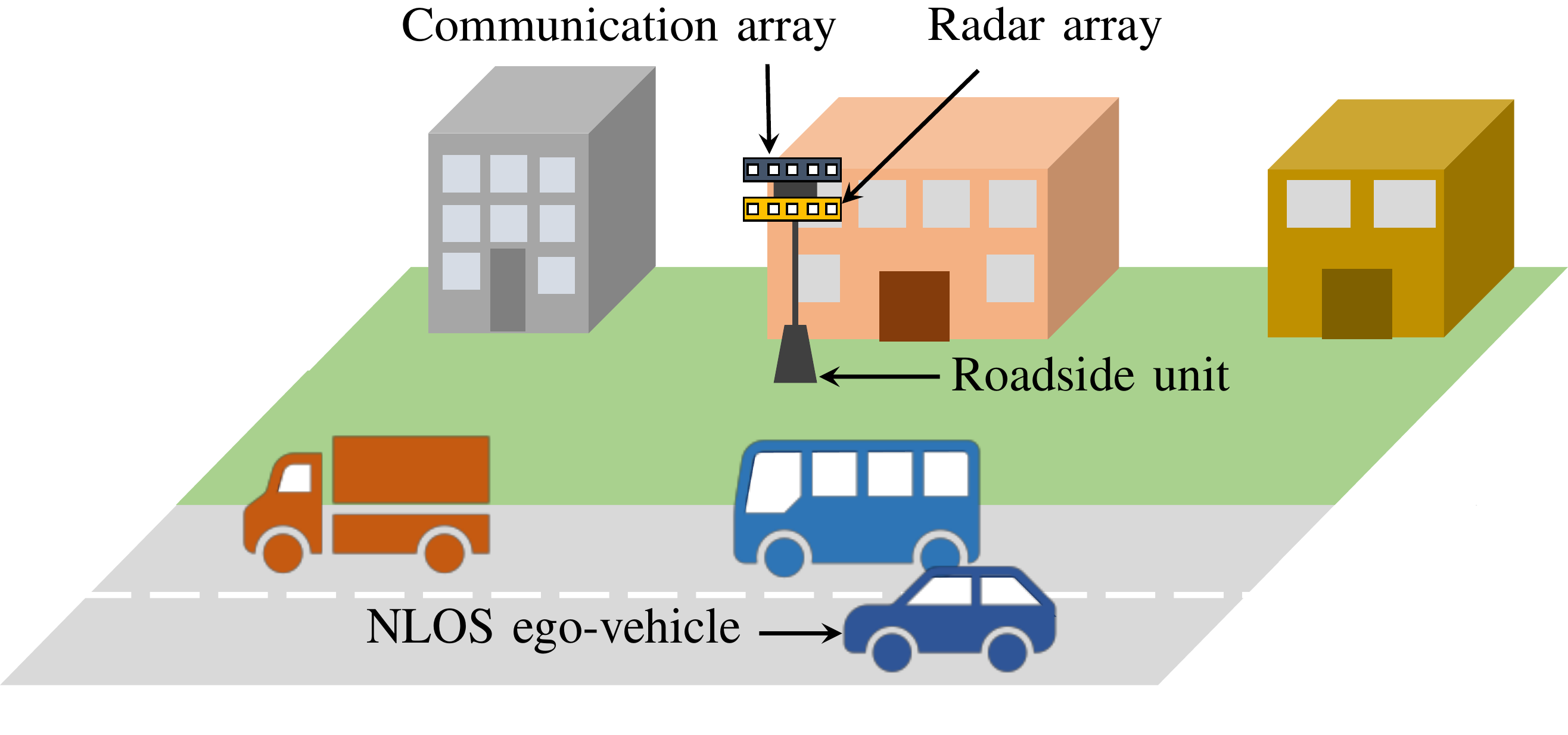}
\caption{The \ac{V2I} communication setup with the \ac{RSU} equipped with a communication and a radar array.}
\label{fig:V2Isetup}
\end{figure}

\begin{figure}[h!]
\centering
\includegraphics[width=0.45\textwidth]{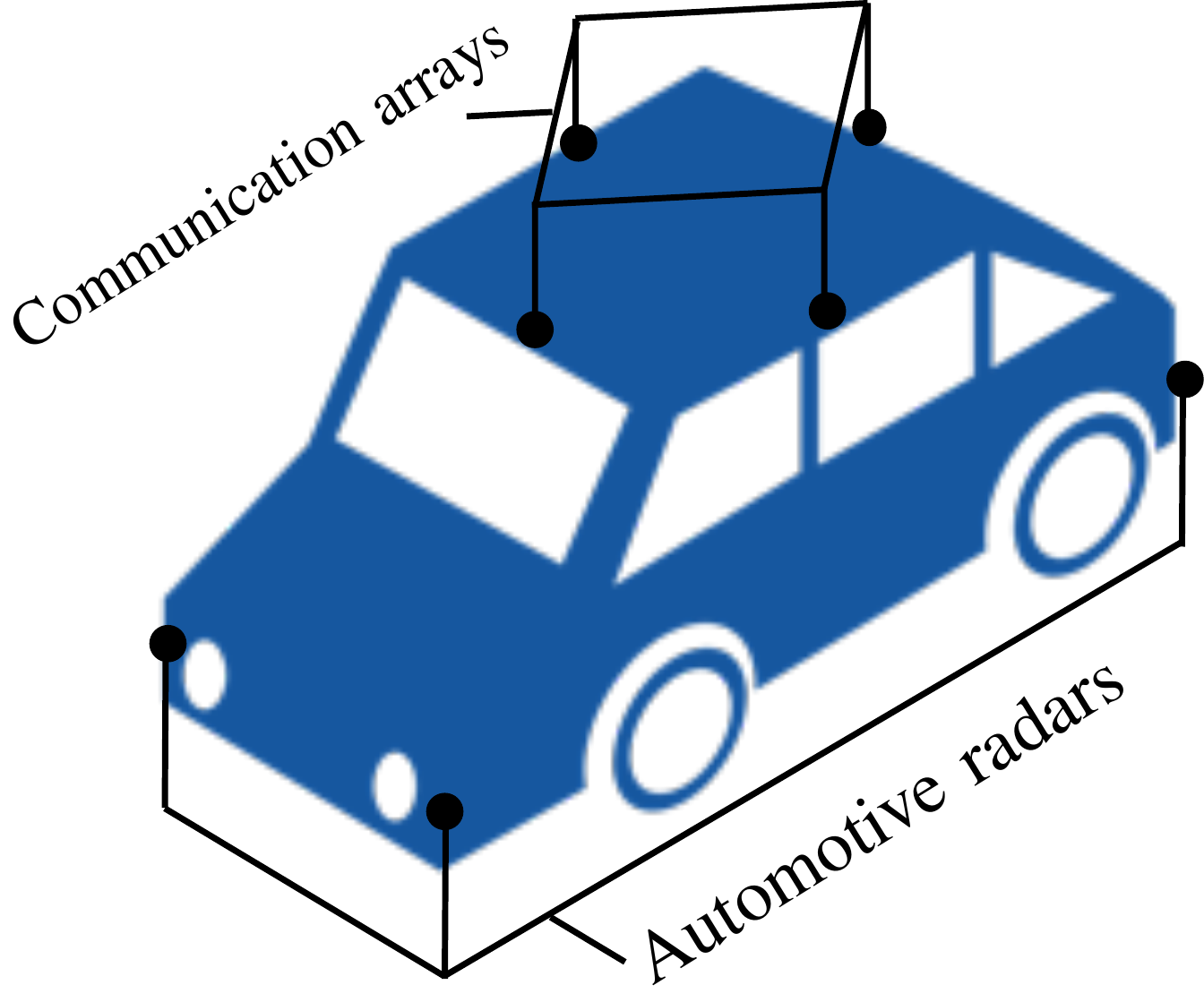}
\caption{The ego-vehicle with multiple communication arrays and multiple radars.}
\label{fig:egoveh}
\end{figure}

We consider the \ac{V2I} communication system as shown in Fig.~\ref{fig:V2Isetup}. The \ac{RSU} is equipped with a communication array and a passive radar array. The communication and radar arrays are collocated and horizontally aligned. The ego-vehicle on the road -- as shown in Fig.~\ref{fig:egoveh} -- is equipped with multiple communication arrays as proposed in \ac{3GPP}~\cite{3GPP37885}. The vehicle is also equipped with multiple \acp{MRR} (e.g., as used in Audi A8~\cite{Padgett2017Heres}). Note that the radars and the communication arrays on the vehicle are not collocated. Our objective is to tap the radar transmissions at the \ac{RSU} to obtain the radar spatial covariance. Subsequently, we seek to use this spatial covariance to configure the mmWave communication link. The developments in this work assume \acp{ULA} for communication between the \ac{RSU} and the vehicle. The radars at the vehicle are single antenna, whereas the \ac{RSU} has a passive \ac{ULA} to tap the radar transmissions. With suitable modifications, the strategies proposed in this work can be extended to other array geometries. 

\subsection{Communication system model}\label{sec:comm}

\begin{figure*}[h!]
\centering
\includegraphics[width=\textwidth]{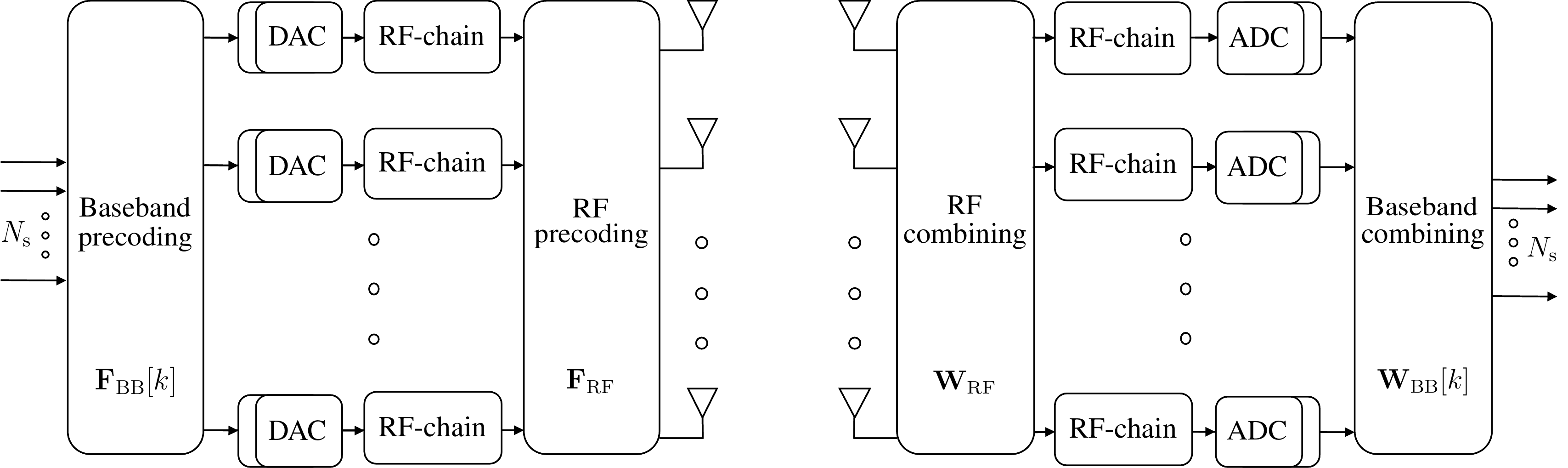}
\caption{The mmWave communication system with hybrid analog-digital precoding and combining.}
\label{fig:mmWavesystem}
\end{figure*}

The mmWave communication system is shown in Fig.~\ref{fig:mmWavesystem}. The \ac{RSU} communication array has $\NRSU$ antennas and $\MRSU\leq\NRSU$ RF-chains. The vehicle has $A$ communication antenna arrays, each with $\NV$ antenna elements and $\MV\leq\NV$ RF-chains. We assume that $\Ns\leq\min\{\MRSU,\MV\}$ data-streams are transmitted. For communication, we consider an \acs{OFDM} system with $K$ sub-carriers. The transmission symbols on sub-carrier $k$ are $\bsfs[k]\in\bbC^{\Ns\times1}$, that follow $\bbE[\bsfs[k]\bsfs^\ast[k]]=\frac{P_\rmc}{K\Ns}\bI_{\Ns}$, where $P_\rmc$ is the total average transmitted power. Let $\bFBB[k]\in\bbC^{\MRSU\times\Ns}$ be a baseband-precoder  and $\bFRF\in\bbC^{\NRSU\times\MRSU}$ be an RF-precoder, then we use $\bF[k]=\bFRF\bFBB[k]\in\bbC^{\NRSU\times\Ns}$ to denote the hybrid precoder on sub-carrier $k$. The RF-precoder is implemented in the time-domain and is common to all sub-carriers. We assume that the RF-precoder is implemented using quantized phase-shifters with a finite set of possible values, i.e., $[\bFRF]_{i,j}=\frac{1}{\sqrt{\NRSU}}e^{\compj \zeta_{i,j}}$, where $\zeta_{i,j}$ is the quantized phase. The precoders satisfy the total power constraint $\sum_{k=1}^{K}\|\bF[k]\|_\rmF^2=K\Ns$. 

We assume perfect time and frequency synchronization at the receiver. Further, let $\bWBB^{(a)}[k]\in\bbC^{\MV\times\Ns}$ be a baseband-combiner, and $\bWRF^{(a)}\in\bbC^{\NV\times\MV}$ be an RF-combiner, then we use $\bW^{(a)}[k]=\bWRF^{(a)}\bWBB^{(a)}[k]\in\bbC^{\NV\times\Ns}$ to denote the hybrid combiner. For the $a$th array on the vehicle, if $\bsfH^{(a)}[k]$ denotes the frequency-domain $\NV\times\NRSU$ mmWave \acs{MIMO} channel on sub-carrier $k$, then the post-processing received signal on sub-carrier $k$ is
\begin{align}
\bsfy^{(a)}[k]=\bsfW^{(a)\ast}[k]\bsfH^{(a)}[k]\bsfF[k]\bsfs[k]+\bsfW^{(a)\ast}[k]\bsfn^{(a)}[k],
\label{eq:rxpost}
\end{align}
where $\bsfn^{(a)}\sim\cC\cN(\bzero,\sigma_{\bsfn}^2\bI)$ is the additive white Gaussian noise. 

\subsection{Channel model} 
We adopt a wideband geometric channel model with $C$ clusters. Each cluster has a mean time-delay $\tau_c \in \bbR$, mean physical \ac{AoA} and \ac{AoD} $\{\theta_c, \phi_c\} \in [0,2\pi)$. Each cluster contributes $R_c$ rays/paths between the \ac{RSU} and the vehicle, where each ray $r_c\in[R_c]$ has a relative time-delay $\tau_{r_c}$, relative angle shift $\{\vartheta_{r_c},\varphi_{r_c}\}$, and a complex path gain $\alpha_{r_c}$. W use $\baV(\theta)$ and $\baRSU(\phi)$ to denote the antenna array response vectors of the vehicle and the \ac{RSU}, respectively. Let $\Delta$ be the inter-element spacing normalized by the wavelength, then the array response vector of the \ac{RSU} is
\begin{align}
\baRSU(\theta)=[1,e^{\compj 2\pi\Delta\sin(\theta)},\cdots,e^{\compj(\NRSU-1) 2\pi\Delta\sin(\theta)}]^\transp.
\end{align}
The array response vectors of the vehicle arrays are defined in a similar manner. Further, let $p(\tau)$ denote the combined effects of analog filtering and pulse shaping filter evaluated at point $\tau$, and let $\Tc$ be the signaling interval. To write the channel model, we remove the superscript $(a)$ from the channel $\bH$ to lighten the notation with the understanding that the channels between the \ac{RSU} and all vehicle arrays follow the same model. Now, the delay-$d$ \ac{MIMO} channel matrix $\bH[d]$ is~\cite{Alkhateeb2016Frequency}
\begin{align}
\bH[d]=\sum_{c=1}^C \sum_{r_c=1}^{R_c} \alpharc p (d\Tc-\tau_c-\tau_{r_c})\baV(\phi_c+\varphi_{r_c})\baRSU^\ast(\theta_c+\vartheta_{r_c}).
\label{eq:timedomch}
\end{align}
If there are $D$ delay-taps in the channel, the channel at sub-carrier $k$, $\bsfH[k]$ is~\cite{Alkhateeb2016Frequency}
\begin{align}
\bsfH[k]=\sum_{d=0}^{D-1}\bH[d] e^{-\compj \tfrac{2\pi k}{K}d}.
\label{eq:freqdomch}
\end{align}

\subsection{Covariance model}
The \ac{RSU} spatial covariance on sub-carrier $k$ is defined as $\bsfRRSU[k]=\frac{1}{\NV}\bbE[\bsfH^\ast[k]\bsfH[k]]$. For the development of the proposed strategies, we make the typical assumption that covariances across all sub-carriers are identical~\cite{Bjornson2009Exploiting}. With this assumption, we can base our designs on a covariance estimate obtained by averaging the estimated covariances across all the sub-carriers and denoted simply as $\hbsfRRSU=\frac{1}{K}\sum_{k=1}^K \hbsfRRSU [k]$. Note that the same covariance across sub-carriers implies that the time domain channel taps are uncorrelated. In practice, however, the channel delay-taps have some correlation and the spatial covariance matrices on all sub-carriers, though similar, are not identical. Furthermore, we ignore the frequency dependence of the antenna array which leads to beam-squint~\cite{Venugopal2019Optimal}. Thus designing the RF and baseband precoders/combiners for all sub-carriers using an averaged covariance will result in some performance loss.  Note, however, that the analog combiner is designed commonly for all sub-carriers. As such, if the covariance is used only for analog precoder and combiner design, it is appropriate to use averaged covariance. The baseband can then be configured independently for all sub-carriers using an estimate of the low dimensional equivalent channel (which includes the physical channel, and the analog precoder and combiner).
\section{Passive FMCW radar}\label{sec:rad}
\begin{figure}[h!]
\centering
\includegraphics[width=0.7\textwidth]{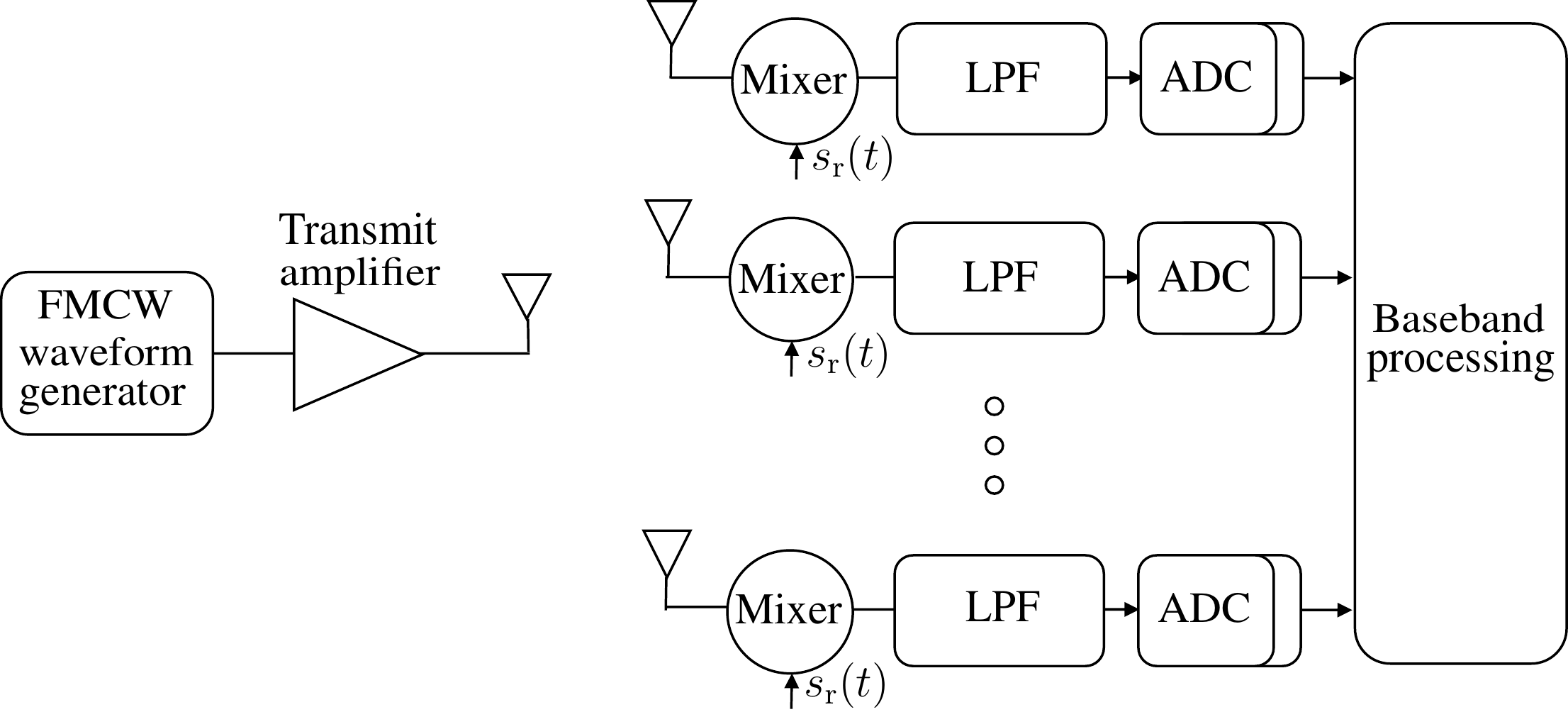}
\caption{The \ac{FMCW} radar system with multiple antennas at the receiver.}
\label{fig:Radarsystem}
\end{figure}

The ego-vehicle shown in Fig.~\ref{fig:egoveh} is equipped with multiple radars. We start by developing the system model for a single radar and later incorporate transmissions from all radars. An \ac{FMCW} radar system is shown in Fig.~\ref{fig:Radarsystem}. For the development, we consider a fully digital receiver at the \ac{RSU}. This assumption can be justified, as, for radar, a fully digital receiver can be emulated by a switching network and a few RF-chains. Specifically, measurements from only a few antennas are collected at a given time. These measurements are then combined by correcting for the effects of sequential sampling to mimic a simultaneous measurement from all antennas. The INRAS Radarbook~\cite{INRASRadarbook} is an example of a radar with this architecture. Specifically, two \acp{ADC} are sequentially connected to four antennas. These four sequential measurements (from two antennas each) are then corrected and combined to obtain a received signal from all the eight receive antennas. 

\subsection{Perfect waveform knowledge at the receiver}
The \ac{FMCW} signals are transmitted in chirps. Let $T_\rmp$ be the chirp duration, and let $B_\rmr$ be the radar bandwidth, then $\beta=\frac{B_\rmr}{T_\rmp}$ is the chirp rate~\cite{Suleymanov2016Design,Hinz2011MIMO}. If we let $f_\rmr$ denote the initial frequency of the radar, then the transmit waveform is~\cite{Suleymanov2016Design,Hinz2011MIMO}
\begin{align}
s_\rmr(t)=\exp\left(\compj 2\pi(f_{\rmr} t + \frac{\beta t^2}{2})\right),~0\leq t < T_\rmp.
\end{align}
The transmit waveform is scaled before transmission so as to have the transmit power $P_\rmr$. With this, the transmitted signal is
\begin{align}    
s(t)&=\sqrt{P_\rmr}s_\rmr(t).
\end{align}

If the radar receiver has $N_\rmr$ antennas, let us denote the received signal on all antennas by a vector $\bx(t)\in \bbC^{N_\rmr}$. The transmitted waveform arrives at the $n$th antenna of the receiver with attenuation $\alpha$ and delay $\tau_n$. The received signal on $n$th antenna is thus
\begin{align}
[\bx(t)]_n=\alpha s(t-\tau_n).
\end{align}
Let $\tau$ be the propagation time of the transmit signal that contains the delay due to distance. Further, let $\tau^\prime_n$ be the additional time-delay of the wave propagating from the reference antenna to the $n$th antenna of the \ac{ULA}. For an incoming signal that has angle $\theta$ relative to the broadside of the array, the delay $\tau^\prime_n$ in a half wavelength spacing \ac{ULA} is
\begin{align}
\tau^\prime_n=\frac{\sin \theta (n-1) }{2 f_\rmr}.
\end{align}
With $\tau$ being the delay due to distance, and $\tau^\prime_n$ being the additional delay, we can write the delay of the transmit waveform on antenna $n$ as $\tau_n=\tau + \tau^\prime_n$~\cite{Katkovnik2002High}.

When the reference signal $s_\rmr(t)$ is known at the receiver, the echo signal $[\bx(t)]_n$ is mixed with the reference signal. The resulting signal is passed through a \ac{LPF} to remove the high frequency components. The noiseless version of mixed and filtered signal is
\begin{align}
[\by(t)]_n= [\bx^\ast(t)]_n s_\rmr(t)=\sqrt{P_\rmr}\alpha \exp\left(\compj 2\pi (f_\rmr \tau_n -\frac{\beta \tau_n^2}{2} + \beta \tau_n t)\right).
\end{align}
The term $\beta\tau_n$ is the constant frequency of the signal (called beat-frequency) that is used to estimate the range of the target. Further, $2\pi \fr \tau_n - \pi \beta \tau_n^2$ is the constant phase of the signal. The variation of this phase across the chirps is used to estimate the Doppler. Let us collect the $I$ samples of the signal in a matrix $\bY\in\bbC^{N_\rmr \times I}$. If $i\in\{1,2,\cdots,I\}$ denotes the sample index, and $\Tr$ denotes the sampling time, then the $i$th sample on the $n$th antenna is
\begin{align}
[\bY]_{n,i}=\sqrt{P_\rmr}\alpha \exp\left(\compj 2\pi (\fr \tau_n -\frac{\beta \tau_n^2}{2} + \beta \tau_n i \Tr)\right).
\label{eq:radarrx}
\end{align}
Note that, so far we have considered a single radar at the vehicle and a single reflector. As discussed previously, the vehicle is equipped with multiple radars. We assume that all the radars transmit at the same time. This is feasible as the radars mounted on a vehicle have exclusive field-of-view, and they do not need to be separated in time to avoid interference. We introduce superscripts to denote the radar number and the reflector number, i.e., the signal received from the $j$th radar and the $l$th reflector is $\bY^{j,l}$. Further, let $\bN$ be the additive white Gaussian noise with entries $\cC\cN(0,\sigma_{n}^2)$. Then, we can write the superimposed radar received signal $\bY_\rmr$ as
\begin{align}
\bY_\rmr=\sum_{j=1}^{J} \sum_{l=1} ^{L} \bY^{j,l}+ \bN.
\label{eq:radarrxfin}
\end{align}

In this work, our objective is to configure the downlink of the mmWave communication system based on the spatial covariance matrix constructed from~\eqref{eq:radarrxfin}. There are, however, two obstacles in this pursuit. First, we have assumed that the reference signal $s_\rmr(t)$ is known at the receiver. In a typical passive radar, a static source illuminates the environment. The receiver has a dedicated channel through which the reference waveform is monitored and sampled. In our application, however, the source is mobile, and it is not possible to have access to the reference signal of the source. The second obstacle is that angular information retrieved from radar covariance has a bias~\cite{Kim2018Joint}. The impact of this bias is significant for a system with a large number of antennas. Specifically, due to a narrow beamwidth in a large antenna system, small pointing errors can result in a significant loss in the link budget. In the following two sub-sections, we address these two challenges.    

\subsection{Proposed radar receiver}\label{sec:radrx}
The reference signal and the transmission time is required to estimate the range (through beat-frequency) and Doppler (through phase variations across chirps). In our application, however, we are interested in the spatial covariance of the radar received signal $\bR_\rmr\in\bbC^{N_\rmr \times N_\rmr}$. The spatial covariance matrix, of the radar received signal (for a single target), can be estimated from the received signal $\bY$ in~\eqref{eq:radarrx} as
\begin{align}
\bR=\frac{1}{I}\bY\bY^\ast.
\end{align}
If we define $\Delta\tau_{qp}=\tau^\prime_q - \tau^\prime_p$, then ignoring the contribution of noise, the covariance between the signal received on the $q$th antenna and the $p$th antenna with perfect reference signal knowledge is
\begin{align}
[\bR_\rmr]_{q,p}=\frac{1}{I}\sum_{i=1}^I \exp\left(\compj 2\pi \Big(f_c \Delta\tau_{qp} -\frac{\beta}{2} (2\tau +\tau^\prime_q + \tau^\prime_p)\Delta\tau_{qp} +\beta i \Tr \Delta\tau_{qp}\Big)\right).
\label{eq:perfcorr}
\end{align}

As we are only interested in the spatial covariance, we propose a simple radar processing chain that does not require the reference signal and achieves the same covariance as in~\eqref{eq:perfcorr}. To this end, let $\Delf$ be the frequency offset between the clock at the \ac{RSU} and the vehicle, and $\epsilon$ be the phase-offset. Then, in the absence of the reference signal, we mix the received signal with $\check s_\rmr(t)=\exp(\compj \{2\pi (\fr+\Delf)t+\epsilon\})$. The received signal after mixing (with $\check s_\rmr(t)$) and filtering is
\begin{align}
[\check\by(t)]_n= [\bx^\ast(t)]_n \check s_\rmr(t)=\sqrt{P_\rmr} \alpha \exp\left(\compj \Big(2\pi \big(\fr \tau_n + \Delf t -\frac{\beta (t-\tau_n)^2}{2}\big) +\epsilon\Big)\right).
\end{align}

With the proposed architecture, the frequency of the received signal $[\check\by(t)]_n$ is $\beta\tau_n + \Delf - \frac{\beta t}{2}$, which is random (due to $\Delf$) and time-varying (due to $\frac{\beta t}{2}$). Thus, the range of the target cannot be estimated by the proposed simplified receiver architecture. Also the phase of the received signal, i.e., $2\pi\fr\tau_n- \pi\beta \tau_n^2 + \epsilon$, is random and varies from one chirp to another (as $\epsilon$ varies from one chirp to another). Thus, the Doppler of the target cannot be estimated either by the proposed simplified receiver architecture. Note that, the $i$th sample of the received signal on antenna $n$ (collected in a matrix $\check\bY\in\bbC^{N_\rmr\times I}$) is
\begin{align}
[\check\bY]_{n,i}=\sqrt{P_\rmr}\alpha \exp\left(\compj 2\pi (\fr \tau_n +\Delf i \Tr - \frac{\beta \tau_n^2}{2} - \frac{\beta i^2 \Tr^2}{2} + \beta \tau_n i \Tr + \epsilon )\right),
\label{eq:radarrxsim}
\end{align}
and the spatial covariance matrix based on~\eqref{eq:radarrxsim} is
\begin{align}
\check \bR_\rmr=\frac{1}{I}\check\bY \check\bY^\ast.
\end{align}

It is easy to show that $[\check\bR_\rmr]_{q,p}=[\bR]_{q,p}$. This observation allows us to circumvent the requirement of the reference signal in our application without any loss in terms of spatial information. Note that, we have shown that the simplified architecture has the same spatial covariance as with perfect waveform knowledge in a single target scenario. This choice was made for ease of exposition, and similarly, it can be shown that the spatial covariance is also the same for the multiple target case. Similar to~\eqref{eq:radarrxfin}, we define $\check\bY^{j,l}$ as the signal received from the $j$th radar and the $l$th reflector in the simplified architecture. Then, the superimposed signal for all the targets and all the radar transmitters is
\begin{align}
\check\bY_\rmr=\sum_{j=1}^{J} \sum_{l=1} ^{L} \check\bY^{j,l}+ \check\bN.
\label{eq:radarrxsimfin}
\end{align}

\subsection{Radar bias correction}\label{sec:biascorr}
In~\cite{Kim2018Joint}, it was shown that the angle estimation based on \ac{FMCW} radar is biased. Specifically, for a single point target at angle $\theta$, the true and the estimated angles are related by~\cite{Kim2018Joint}
\begin{align}
\sin \hat\theta=\left(1+\frac{B_\rmr}{2 \fr}\right)\sin\theta,
\label{eq:radarbias}
\end{align}    
in a noiseless scenario. For a system with a large number of antennas - where the beams are narrow - this multiplicative bias can have a significant impact on the performance. As an example, note that the first-null of a \ac{ULA} with $N$ antennas is $2/N$ away from the main lobe (say centered at $\sin(\theta)$)~\cite{Orfanidis2016Electromagnetic}. For $N=256$, we get $2/N=7.8\times 10^{-3}$. Similarly, for $\fr=\SI{76}{\giga\hertz}$, and $B_\rmr=\SI{1.2}{\giga\hertz}$ (i.e., a fraction of bandwidth available in $\SI{76}{\giga\hertz}$ band), we have $\frac{B_\rmr}{2 f_\rmr}=7.9\times 10^{-3}$. Thus, even in the noiseless case, the beamforming based on the biased estimate can imply a null in the direction of the true angle. 

A similar error/mismatch appears in \ac{FDD} systems. Assuming perfect angular reciprocity is reasonable in \ac{FDD} systems, as the \ac{UL} and \ac{DL} are only separated by hundreds of $\SI{}{\mega\hertz}$. Therefore, the differences in the \ac{UL} and \ac{DL} covariance come only from the array response in the \ac{UL} and \ac{DL}. Assume that the array is a \ac{ULA} with inter-element spacing set to half the \ac{DL} wavelength. Further, let $\lambda_{\rm DL}$ and $\lambda_{\rm UL}$ denote the \ac{DL} and \ac{UL} wavelength, then the true and estimated angles in the \ac{UL} will be related by
\begin{align}
\sin \hat\theta=\frac{\lambda_{\rm DL}}{\lambda_{\rm UL}}\sin \theta.
\end{align}    

Beamforming based on the biased angle is sub-optimal, especially for a large number of antennas as discussed earlier. Further, note that the biased angle information is recovered from covariance matrices. Covariance matrices, however, can be used for other purposes, e.g., precoder/combiner design based on their singular vectors~\cite{Park2019Spatial,Ali2018Spatial}. As such, it is of interest to correct the covariance matrices directly rather than the angles estimated from the covariance matrices. This problem has been considered in the past for \ac{FDD} systems and several strategies have been proposed, e.g.,~\cite{Aste1998Downlink,Liang2001Downlink,Jordan2009Conversion,Decurninge2015Channel,Khalilsarai2018FDD}. Noting that the bias in the \ac{FMCW} radar has the same form as of that in \ac{FDD}, we use a covariance correction strategy, i.e.,~\cite{Jordan2009Conversion} in this work. 

The strategy~\cite{Jordan2009Conversion} is based on the interpolation of the covariance for correction. The sampled radar covariance matrix $\check \bR_\rmr$ is not necessarily Toeplitz. Therefore, to improve the estimate, we project $\check \bR_\rmr$ to the Toeplitz, Hermitian, positive semi-definite cone $\bT_{+}^N$, i.e.,
\begin{align}
\hat \bR_\rmr=     \underset{\bX\in \bT_{+}^N    }{\arg\min} \|\bX-(\check \bR_\rmr - \sigma_n^2 \bI)\|_\rmF.
\label{eq:Toeplitzproj}
\end{align}
As $\hat \bR_\rmr$ is a Toeplitz Hermitian matrix, it is fully described by its first column, which we denote as $\hat \br$.

\begin{figure}[h!]
\centering
    \centering
    \begin{subfigure}[t]{0.45\textwidth}
        \centering
        \includegraphics[width=1\textwidth]{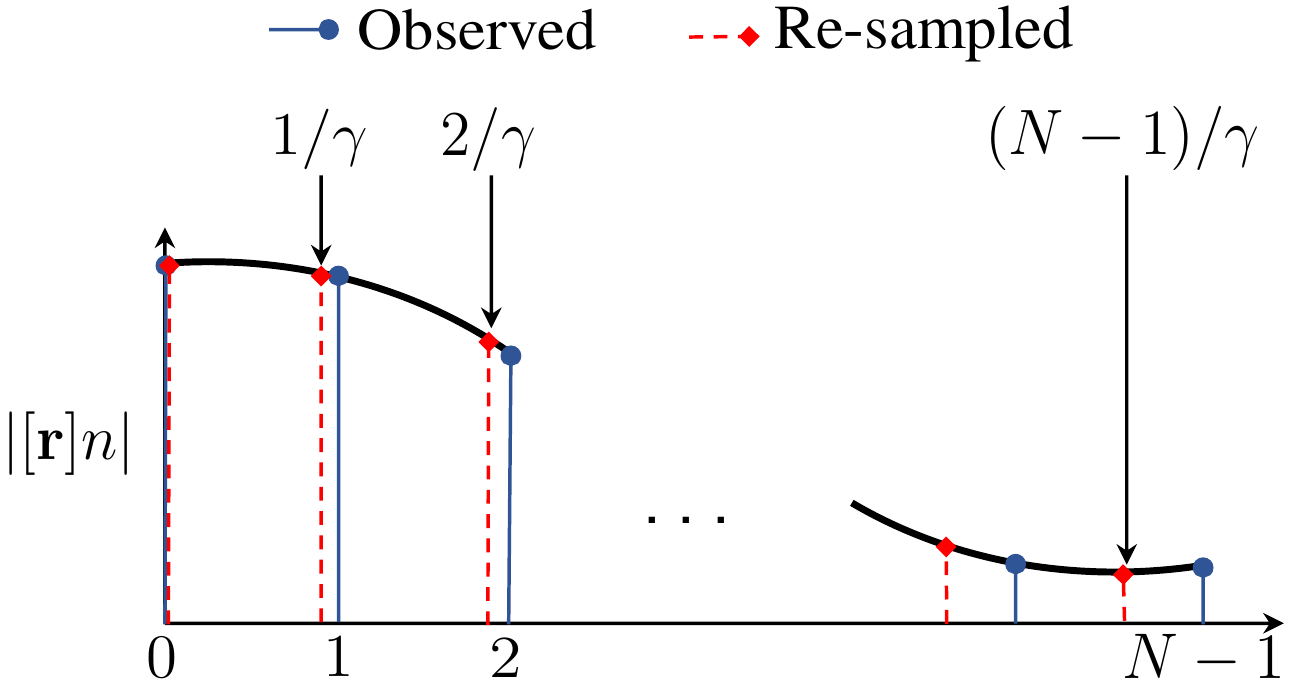}
        \caption{Observed and resampled magnitude of the covariance vector $\br$ for covariance correction.}
        \label{fig:Mag_resample}
    \end{subfigure}\\
    \begin{subfigure}[t]{0.45\textwidth}
        \centering
        \includegraphics[width=1\textwidth]{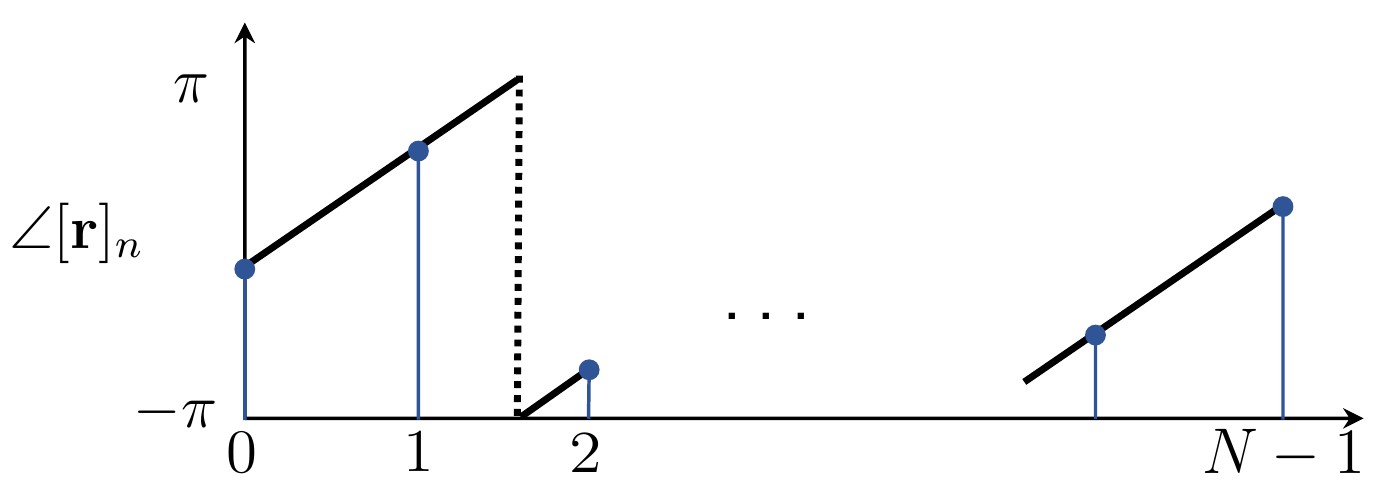}
        \caption{Observed wrapped phase of the covariance vector $\br$.}
        \label{fig:Observed_phase}
    \end{subfigure}\\
        \begin{subfigure}[t]{0.45\textwidth}
        \centering
        \includegraphics[width=1\textwidth]{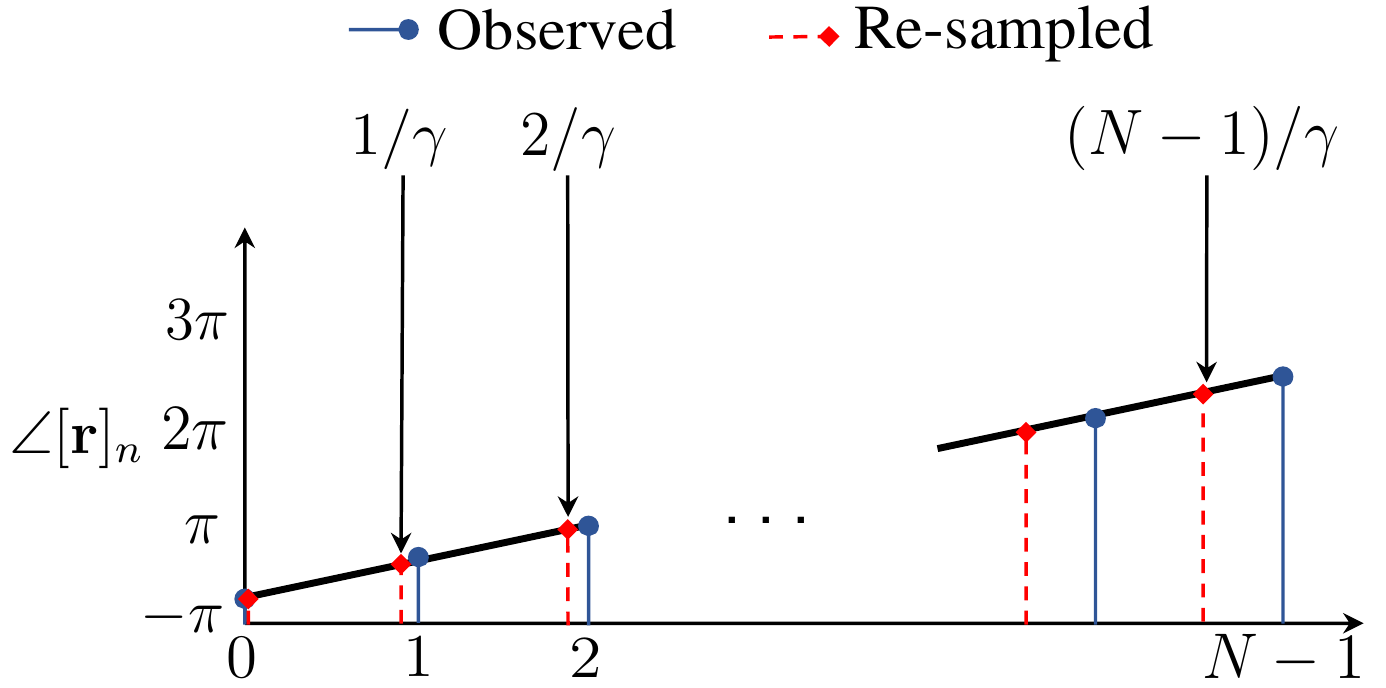}
        \caption{Unwrapped and resampled phase of the covariance vector $\br$.}
        \label{fig:Phase_resample}
    \end{subfigure}
  \hfill
\caption{The covariance correction strategy that (i) resamples the covariance vector magnitude, (ii) unwraps and resamples the phase of the covariance vector, and (iii) uses Toeplitz completion of the resampled covariance vector to obtain the corrected covariance.}
\label{fig:Resampling}
\end{figure}

For simplicity, let us denote the multiplication constant in~\eqref{eq:radarbias} as $\gamma=(1+\frac{B_\rmr}{2 \fr})$. Note that the vector $\hat\br$ contains the samples of the covariance at $n\in \{0,1,\cdots,N-1\}$. For correction, we need the samples at points $n/ \gamma~\forall n\in \{0,1,\cdots,N-1\}$. This can be achieved by interpolation (note that as $\gamma>1$, there is no need for extrapolation). As the vector $\hat\br$ is complex, we interpolate the magnitude and phase separately. The magnitude of $\hat\br$ is smooth, and \emph{spline} interpolation followed by resampling will provide good performance as shown in Fig.~\ref{fig:Mag_resample}. The phase of $\hat\br$ can be unambiguously determined only in the interval $(-\pi,\pi]$ as shown in Fig.~\ref{fig:Observed_phase}. That said, the phase changes slowly with $n$, and hence the jumps of $2\pi$ can be observed. For the interpolation of phase, the actual phase needs to be reconstructed. Hence, the observed phase is \emph{unwrapped} first as shown in Fig~\ref{fig:Phase_resample}, and then spline interpolation followed by resampling is used. We can then obtain the corrected covariance vector $\hat\br_\rmc$ by combining the resampled magnitude and phase. Finally, the corrected covariance matrix $\hat\bR_\rmc$ is obtained by Toeplitz completion $\hat\bR_\rmc=\cT(\hat\br_\rmc)$.
\section{Similarity metric to measure spatial congruence}\label{sec:congruence}
\begin{figure}[h!]
\centering
    \centering
    \begin{subfigure}[t]{0.25\textwidth}
        \centering
        \includegraphics[width=1\textwidth]{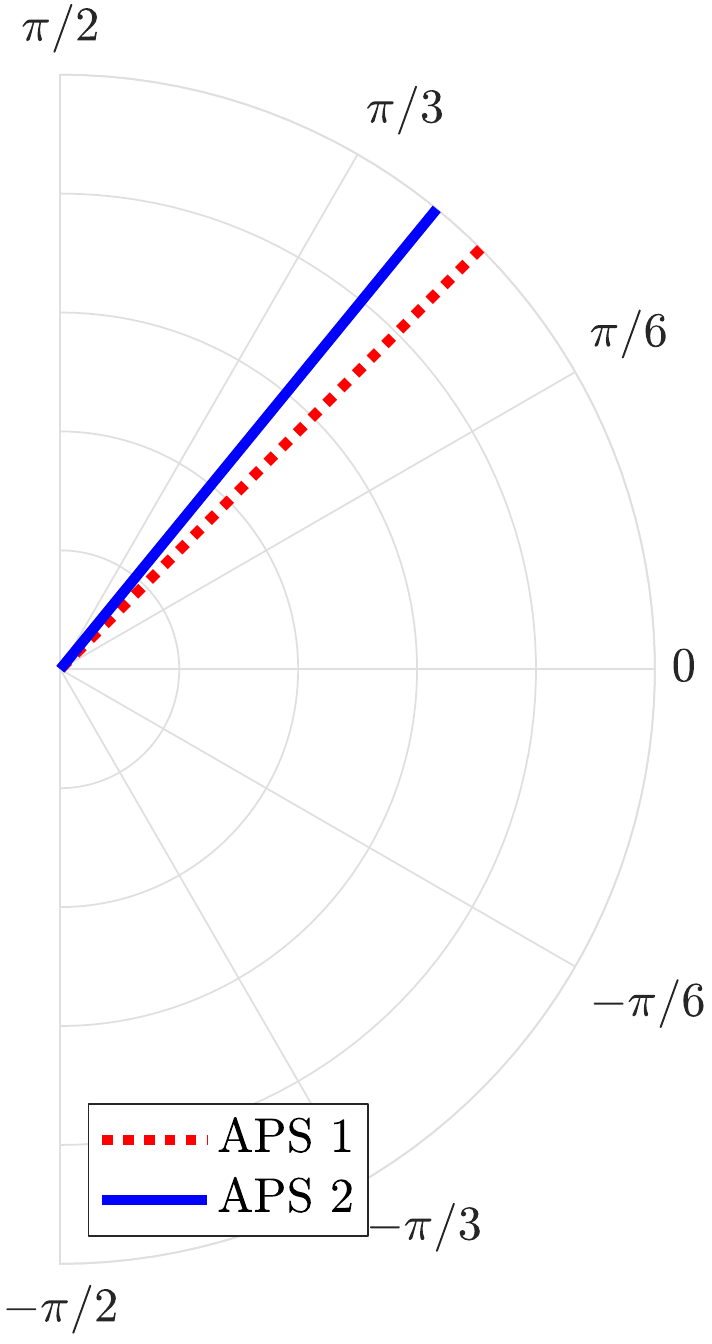}
        \caption{Two over-the-air APS.}
        \label{fig:APS}
    \end{subfigure}%
    \begin{subfigure}[t]{0.25\textwidth}
        \centering
        \includegraphics[width=1\textwidth]{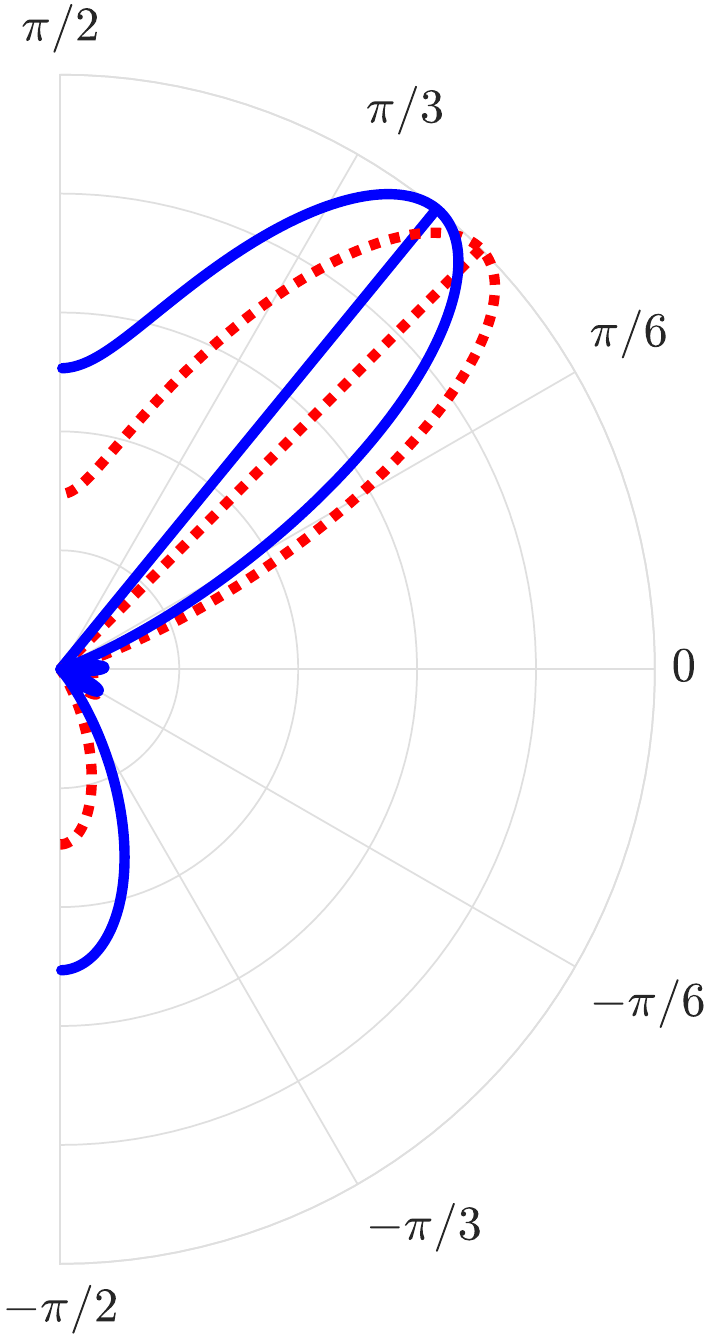}
        \caption{Overlaid beams for a $4$ antenna array.}
        \label{fig:APS4ant}
    \end{subfigure}
        \begin{subfigure}[t]{0.25\textwidth}
        \centering
        \includegraphics[width=1\textwidth]{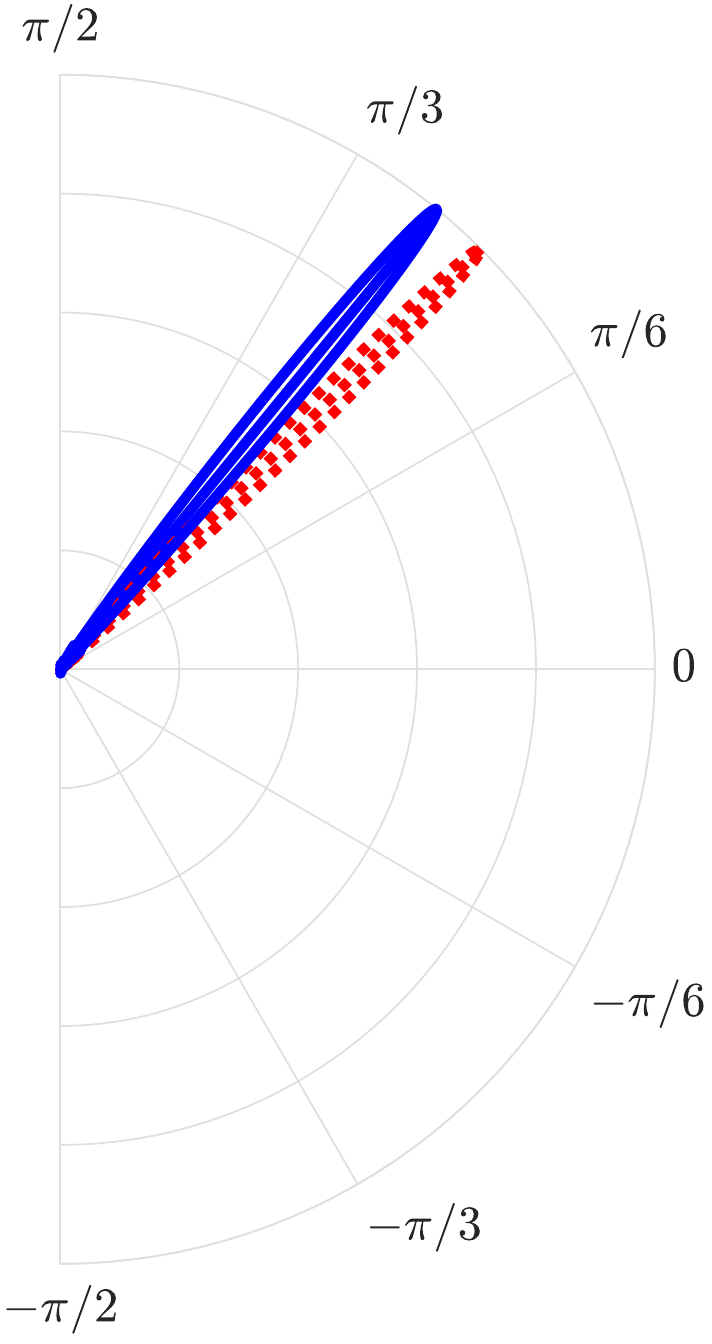}
        \caption{Overlaid beams for a $32$ antenna array.}
        \label{fig:APS32ant}
    \end{subfigure}
  \hfill
\caption{Over the air azimuth power spectra and observed power spectra through arrays of $4$ antenna elements and $32$ antenna elements.}
\label{fig:APSandbeams}
\end{figure}

In this work, we are proposing to use the information retrieved from radar to configure the mmWave communication link. This strategy, however, will be only useful if the spatial characteristics of the radar and the communication channel are congruent. Roughly speaking, we are interested in the similarity of \ac{APS} of the radar received signals and communication channels. In our application, the differences in radar and communication \ac{APS} will stem from: (i) different operating frequencies of radar and communication, and (ii) different locations and \acp{FoV} of communication and radars on the vehicle - hence the different probability of blockage. We, however, need the radar and communication \ac{APS} to be as similar as possible, and a way of quantifying this similarity. Furthermore, we need a similarity metric that is meaningful from the communication system point of view, i.e., the similarity metric should have a transparent connection with a communication system metric, e.g., rate. Such a similarity metric to compare the \ac{APS} will be useful beyond our current application. The metric will be useful, for example, to assess the accuracy of the angular reciprocity assumption in \ac{FDD}. The proposed similarity metric can also be used to validate the assumption that sub-6 GHz and mmWave channels are spatially similar. This assumption is used in a recent line of work~\cite{Ali2018Millimeter,Ali2018Spatial} to reduce the mmWave training overhead using sub-6 GHz information. 

To assess the similarity, one possibility is to compare over-the-air \ac{APS} as shown in Fig.~\ref{fig:APS}. Note that over-the-air \ac{APS} is system independent, i.e., it does not take into consideration how many antennas are used in a system. Let us consider a toy example, where we are interested in measuring the similarity of the \ac{APS}1 (shown with dotted line)  and \ac{APS}2 (shown with solid line). More specifically, consider that we observe \ac{APS}1 (in our case through radar), whereas the actual spectrum is \ac{APS}2 (in our case the \ac{APS} of communication). Now, we consider two cases; (i) a $4$ antenna system, and (ii) a $32$ antenna system. The beam-patterns of $4$ element \acp{ULA} pointing in the directions of the spectra are shown in Fig.~\ref{fig:APS4ant}. In this case, the information provided by \ac{APS}1 is useful for beamforming on APS2. This is because the beam-pattern has a significant gain in the direction of \ac{APS}2. Now for $32$ antennas (the beam-patterns for $32$-element \acp{ULA} are shown in Fig.~\ref{fig:APS32ant}), the information provided by \ac{APS}1 is not particularly useful for beamforming on \ac{APS}2. This is because the beam-pattern directed towards \ac{APS}1 does not have a high gain in the direction of \ac{APS}2. Thus, a meaningful measure of similarity needs to take the system dimension, i.e., the number of antennas, into consideration. 

To define the similarity metric, assume that we want to measure the similarity of two $N$ point spectra $\bd_1$ and $\bd_2$. Consider the index set $\cI_1$ ($\cI_2$) of cardinality $L\leq N$ that contains indices of $L$ largest entries of $\bd_1$ ($\bd_2$). Then, we define a similarity metric
\begin{align}
\rmS_{1 \rightarrow 2}(L,N)= \frac{\sum_{i\in \cI_{1}} \bd_2[i]}{\sum_{i\in \cI_{2}} \bd_2[i]}.
\label{eq:simmet1}
\end{align}

\begin{figure}[h!]
\centering
\includegraphics[width=0.7\textwidth]{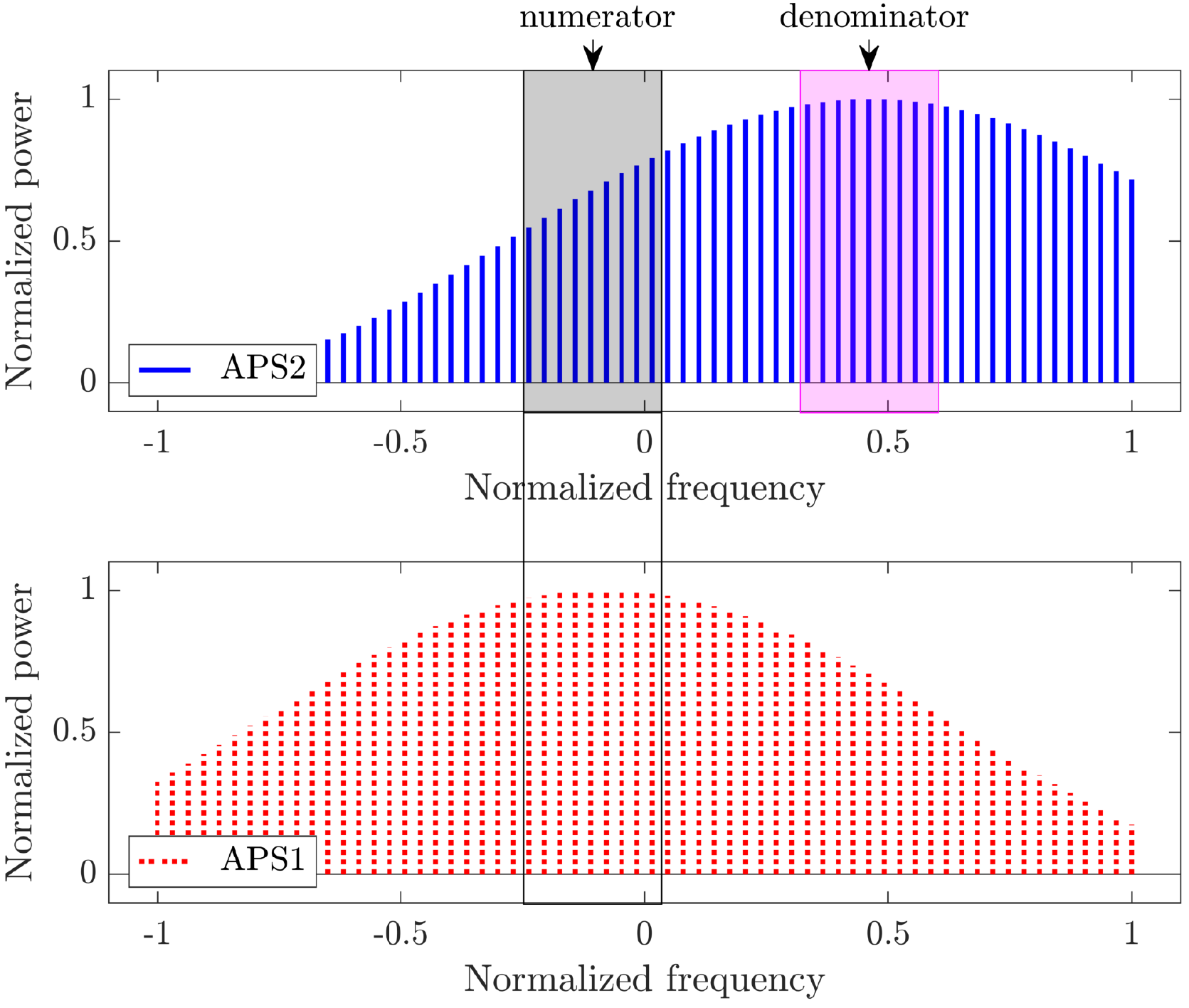}
\caption{Intuitive explanation of the similarity metric~\eqref{eq:simmet1}. The denominator is the sum of a few strongest component of \ac{APS}$2$, whereas the numerator is the sum of a few components of \ac{APS}$2$, corresponding to the strongest components of \ac{APS}$1$.}
\label{fig:Simmeasint}
\end{figure}

To explain the meaning of the metric, we use the help of Fig.~\ref{fig:Simmeasint}. 
In the denominator, we have the sum of $L$ largest spectral components of $\bd_2$, whereas in the numerator we have the $L$ components of $\bd_2$ that correspond to the $L$ largest spectral components of $\bd_1$. It is clear that $0\leq \rmS_{1 \rightarrow 2}(L,N) \leq 1$. For given spectra, as we sum over the $L$ largest spectral components, $\rmS_{1 \rightarrow 2}(L,N)$ will generally increase with $L$ (for fixed $N$), and $\rmS_{1 \rightarrow 2}=1$ for $L=N$. Also, $\rmS_{1 \rightarrow 2}(L,N)$ will generally decrease with $N$ (for fixed $L$) due to narrower beams. Also note that the metric is asymmetric, i.e., it is not necessary that $\rmS_{1 \rightarrow 2}(L,N)= \rmS_{2 \rightarrow 1}(L,N)$. Finally, note that we have used the notation $\rmS_{1 \rightarrow 2}(L,N)$ to include all the relevant parameters. When there is no ambiguity, we will simply use $\rmS$ to denote the similarity metric. Further, note that from a system's point of view $N$ is related to the number of antennas, and $L$ is related to the number of transmitted streams, i.e., $\Ns$.

We now relate the proposed similarity metric to \ac{RPE}~\cite{Park2019Spatial,Haghighatshoar2017Massive}. Consider two channels with spatial covariance matrices $\bR_1$ and $\bR_2$. Further, consider that $\bF_1$ ($\bF_2$) contains the $L$ columns of the \acs{DFT} matrix corresponding to the $L$ largest spectral components of the channel $1$ ($2$). Then, an alternative way to write the proposed similarity metric~\eqref{eq:simmet1} is
\begin{align}
\rmS=\frac{\tr(\bF_1^\ast\bR_2\bF_1)}{\tr(\bF_2^\ast\bR_2\bF_2)}.
\label{eq:simmet2}
\end{align}
The equivalence between~\eqref{eq:simmet1} and~\eqref{eq:simmet2} becomes clear when we note that the \ac{APS} $\bd$ can be written as $\bd=\diag(\bF^\ast\bR\bF)$. Further, if $\bU_1$ ($\bU_2$) are $L$ singular vector of $\bR_1$ ($\bR_2$) corresponding to $L$ largest singular values, then the \ac{RPE} is defined as \cite{Park2019Spatial,Haghighatshoar2017Massive}
\begin{align}
\rm{RPE}=\frac{\tr(\bU_1^\ast\bR_2\bU_1)}{\tr(\bU_2^\ast\bR_2\bU_2)}.
\label{eq:RPE}
\end{align}
In the special case, where the \acp{AoA} of all the paths fall on the \ac{DFT} grid, it is easy to see that the proposed metric and the \ac{RPE} are the same. This is because, for the on-grid case, vectors of the Fourier basis are valid singular vectors. Therefore $\bU_1$ ($\bU_2$) and $\bF_1$ ($\bF_2$) are the same, and the similarity metric in~\eqref{eq:simmet2} is the same as the \ac{RPE} in~\eqref{eq:RPE}. This also implies that asymptotically (i.e., as $N\rightarrow \infty$), the proposed similarity metric is the same as the \ac{RPE}. This is because one way to interpret the asymptotic case is to have a continuous \ac{DFT} grid and hence all \acp{AoA} fall on-grid. It is, however, difficult to analytically relate the proposed metric to the \ac{RPE} in the general off-grid case. It is possible to study this off-grid scenario by simulations instead.

Finally, note that in~\cite{Park2019Spatial}, the \ac{RPE} was related to the relative rate. To understand the relative rate, consider that true covariance is $\bR_2$ (in our case the communication channel covariance), whereas we have access to $\bR_1$ (in our case through radar). The relative rate is then the ratio of the achievable rate given $\bR_1$ to the achievable rate given $\bR_2$. Particularly, in~\cite{Park2019Spatial}, it was shown that in low \ac{SNR} settings the \ac{RPE} is a good approximation of the relative rate. With this connection and the connection between the similarity metric and RPE, the proposed similarity metric also directly relates to the relative rate. To conclude, we proposed a similarity metric to compare two power spectra that relates directly to the relative rate.    
\section{Simulation results}\label{sec:simres} 
In this section, we provide simulation results to verify the ideas presented in this work. We start by discussing the simulation setup in detail. Then, we present results to show the utility of the bias correction strategy presented in Section~\ref{sec:biascorr}, and to numerically study the relationship between the similarity metric (presented in Section~\ref{sec:congruence}) and the \ac{RPE} for the off-grid scenario. Finally, we present achievable rate results to verify the potential of using passive radar to configure mmWave links.

\begin{figure}[h!]
\centering
    \centering
    \begin{subfigure}[t]{0.45\textwidth}
        \centering
        \includegraphics[width=1\textwidth]{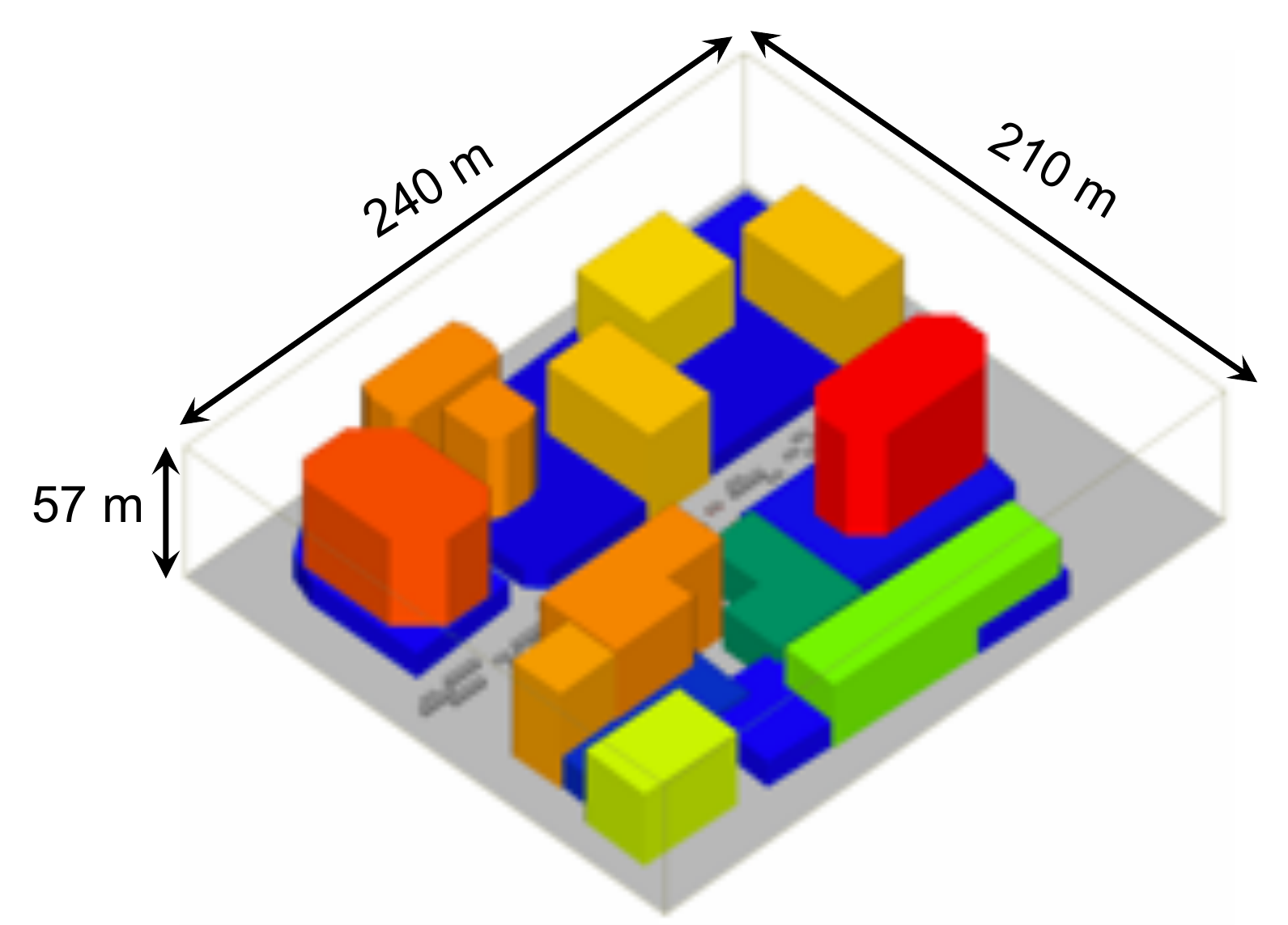}
        \caption{The ray-tracing setup of size $57 \times 240\times 210 \SI{}{\meter}$.}
        \label{fig:WIa}
    \end{subfigure}%
    \begin{subfigure}[t]{0.45\textwidth}
        \centering
        \includegraphics[width=1\textwidth]{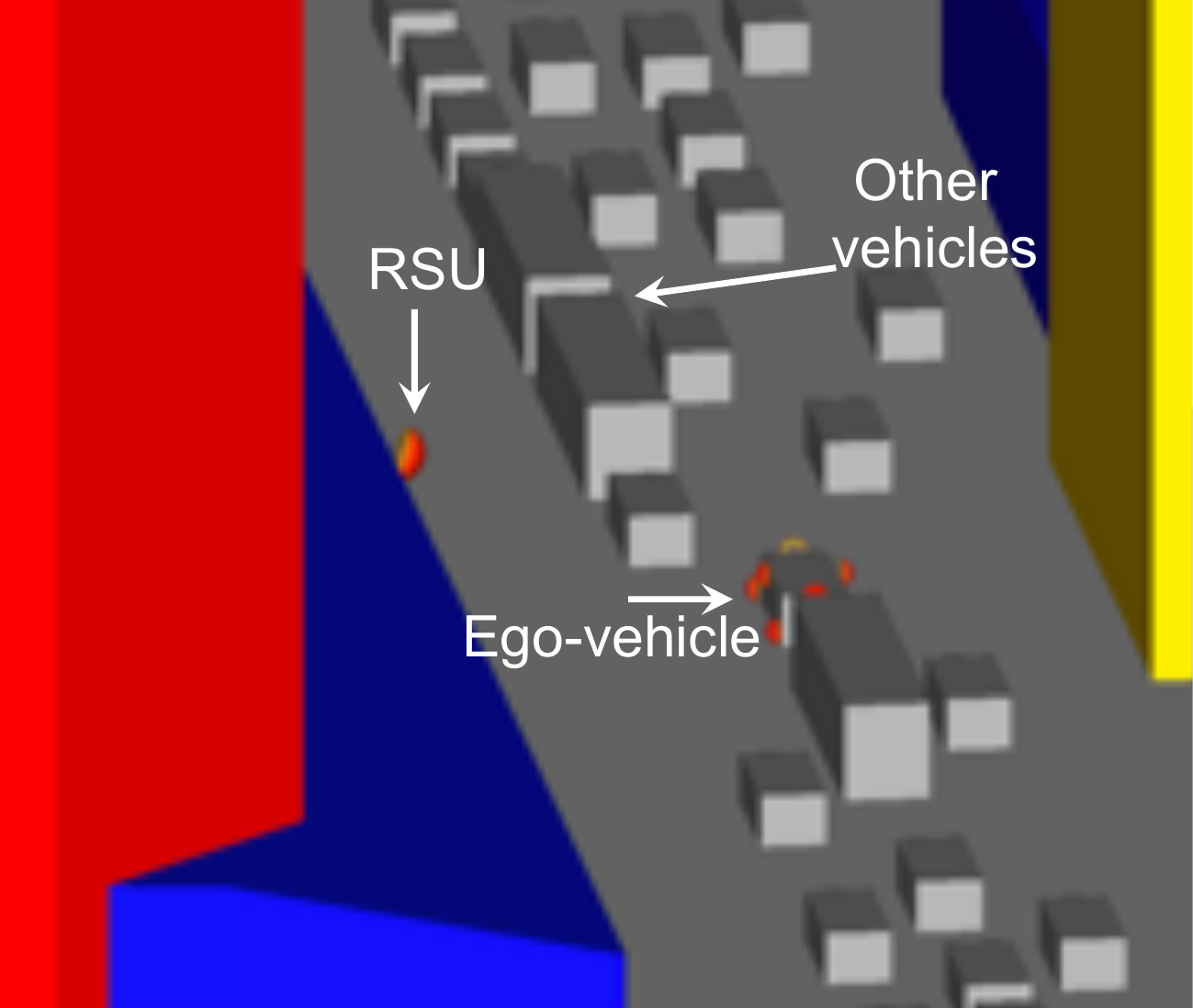}
        \caption{RSU and a randomly selected ego-vehicle.}
        \label{fig:WIb}
    \end{subfigure}
        \begin{subfigure}[t]{0.45\textwidth}
        \centering
        \includegraphics[width=1\textwidth]{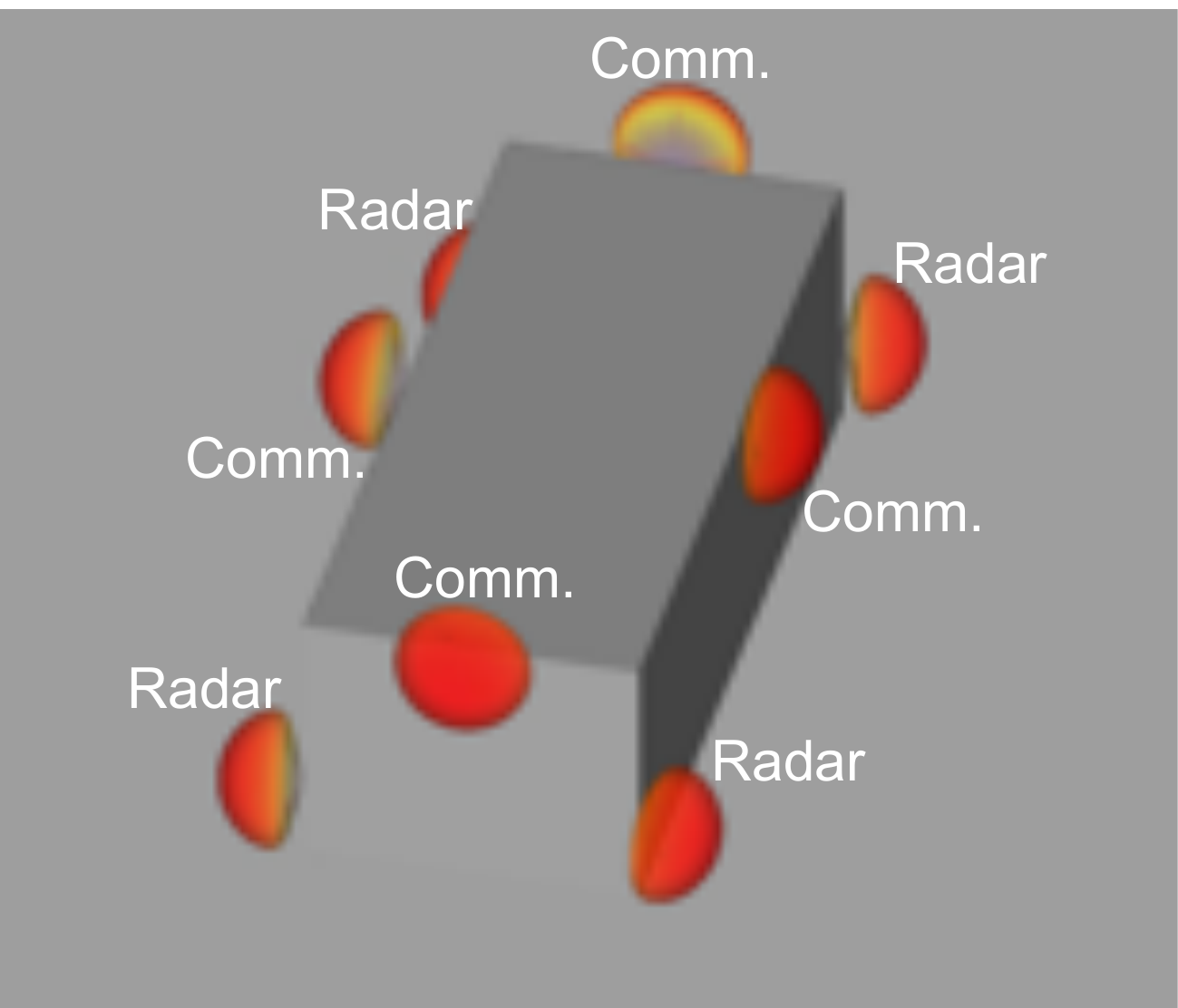}
        \caption{Four communication arrays and four radars on the ego-vehicle.}
        \label{fig:WIc}
    \end{subfigure}
  \hfill
\caption{The ray-tracing setup simulation in Wireless Insite with buildings of various heights, the RSU, and vehicles dropped in four lanes on the road.}
\label{fig:WI}
\end{figure}

\textbf{Material properties for ray-tracing:} For all experiments, we assume that the communication system operates in the $\SI{73}{\giga\hertz}$ band, and the radar operates in the $\SI{76}{\giga\hertz}$ band. We use Wireless Insite~\cite{WI} for ray-tracing simulations. The simulation environment is shown in Fig.~\ref{fig:WIa}. This is an urban environment with buildings on both sides of the road. The color of a building corresponds to its height through a red-green-blue color scale with red representing high and blue representing low. The total size of the ray-tracing setup is around $57 \times 240\times 210 \SI{}{\meter}$. The buildings are made of concrete. The relative permittivity of concrete is $5.31$ and conductivity is $\SI{1.0509}{\siemens\per\meter}$ at $\SI{73}{\giga\hertz}$ (i.e., communication band), and $\SI{1.0858}{\siemens\per\meter}$ at $\SI{76}{\giga\hertz}$ (i.e., radar band)~\cite[Table 3]{ITU2015Effects}. The road surface is made of asphalt with relative permittivity $3.18$ and conductivity $\SI{0.4061}{\siemens\per\meter}$ at $\SI{73}{\giga\hertz}$, and $\SI{0.4227}{\siemens\per\meter}$ at $\SI{76}{\giga\hertz}$~\cite{Li1999Low}. The \ac{RMS} surface roughness for concrete is $\SI{0.2}{\milli\meter}$ and for asphalt is $\SI{0.34}{\milli\meter}$~\cite[Table 1]{Li1999Low}. Wireless Insite also models the diffuse scattering effects. The level of diffuse scattering is controlled using a scattering coefficient with valid values in the range $[0,1]$~\cite{Remcom5G}. We use the scattering coefficient of $0.4$ for concrete~\cite{Remcom5G}. For asphalt, note that the \ac{RMS} surface roughness is higher than concrete, and as diffuse scattering increases with surface roughness~\cite{Pascual-Garcia2016importance}, we choose the scattering coefficient to be $0.5$. Further, some of the diffused power becomes cross-polarized relative to the polarization of the incident ray. In Wireless Insite, this fraction is controlled using the cross-polarization coefficient that has a valid range $[0,0.5]$~\cite{Remcom5G}. We chose the cross-polarization to be half the diffuse scattering coefficient for both concrete and asphalt. Finally, the vehicles on the road are made of metal, i.e., perfect electric conductor.

\textbf{Vehicle size and distribution:} There are two types of vehicles on the road. Vehicles of size $5\times2\times1.6 \SI{}{\meter}$, that represent cars, and vehicles of size $13\times2.6\times 3 \SI{}{\meter}$, that represent trucks~\cite[6.1.2]{3GPP37885}. There are $80\%$ cars and $20\%$ trucks on the road. There are a total of four lanes, each $\SI{3.5}{\meter}$ wide. All the vehicles inside a lane have the same speed. The lane-speeds $s_\ell$ are $60$, $50$, $25$, and $\SI{15}{\kilo\meter\per\hour}$. The fraction of cars and trucks, lane widths, and the lane speeds are the option B for Urban scenarios as proposed by \ac{3GPP} in~\cite[6.1.2]{3GPP37885}.  Let $X$ be an exponential random variable with mean $\mu=s_\ell\times \SI{2}{\second}$ (where $s_\ell$ is in $\SI{}{\meter\per\second}$). Then, the distance between the rear bumper of a vehicle and the front bumper of the following vehicle is $\max(2,x)$~\cite[6.1.2]{3GPP37885}. Note that, the data collected from Wireless Insite is for a time snapshot. As such, the only role of the speed is in calculating inter-vehicle distances. All the results presented in this section are averaged over $1000$ random snapshots, where the vehicles are placed independently in each snapshot according to the mentioned criterion.

\textbf{Antenna locations on vehicle and \ac{RSU}:} The \ac{RSU} (shown in Fig.~\ref{fig:WIb}) has a height of the $\SI{5}{\meter}$~\cite[6.1.4]{3GPP37885}. The radar and communication arrays on the \ac{RSU} are horizontally aligned and are vertically separated by $\SI{10}{\centi\meter}$. The \ac{RSU} arrays are down-tilted so as to face the center of the four lanes. The ego-vehicle has $4$ communication antenna arrays at the height of $\SI{1.6}{\meter}$~\cite[6.1.2]{3GPP37885}, one on each side as shown in Fig.~\ref{fig:WIc}~\cite[Table 6.1.4-9]{3GPP37885}. The radars on the vehicle are placed at the height of $\SI{0.75}{\meter}$. The front radars on the right and left side have $\SI{10}{\degree}$ rotation towards the front. Similarly, the back radars on the right and left side have $\SI{10}{\degree}$ rotation towards the back. The location, height, and the rotations are the numbers chosen to mimic Audi A8 \acp{MRR}~\cite{Padgett2017Heres}.  All antenna elements have $\SI{120}{\degree}$ $\SI{3}{\decibel}$ beamwidth, and $\SI{150}{\degree}$ $\SI{3}{\decibel}$ first null beamwidth (both in E and H-plane). This choice is justified as practical \acp{ULA} also have a \ac{FoV} of around $\SI{120}{\degree}$. We select the ego-vehicle from the vehicles on the road uniformly albeit inside the \ac{FoV} of the \ac{RSU} array.

\textbf{Communication system parameters:} We use the transmit power $P_\rmc=30~\dBm$,  and the bandwidth $B_\rmc=\SI{1}{\giga\hertz}$ for the communication system. We use the raised-cosine filter with a roll-off factor $0.4$ for pulse shaping. Based on the \ac{RMS} delay-spread of the channels obtained through ray-tracing, the bandwidth, and the roll-off factor, the number of time-domain taps required can be calculated to be $D=512$. We use a \ac{CP} of length $D-1$. Further, we choose the number of sub-carriers to be $K=2048$, i.e., the useful part of the \ac{OFDM} symbol is almost $4\times$ the \ac{CP}. The \acp{ULA} used in the communication system have half-wavelength inter-element spacing.

\textbf{Radar parameters:} We consider the chirp period of $T_\rmp=\SI{500}{\micro\second}$, $I=1024$  samples in a chirp, and $128$ chirps for radar processing. The transmit power is $P_\rmr=30~\dBm$ and the bandwidth is $B_\rmr=\SI{1}{\giga\hertz}$. With these parameters the chirp rate is $\beta=\SI{2}{\giga\hertz\per\milli\second}$. For simplified receiver architecture discussed in Section~\ref{sec:radrx}, we select $\Delta f \sim \cU [0,f_{\rm max}]$, with $f_{\rm max}=\SI{3}{\mega\hertz}$, which is around $40$ \ac{ppm} at $\SI{76}{\giga\hertz}$. The phase offset is $\epsilon\sim\cU[0,2\pi]$. The \ac{ULA} used on the \ac{RSU} for radar has half-wavelength inter-element spacing. In all experiments, the number of antenna elements in the radar and communication arrays at the \ac{RSU} are same. 

We modeled the optimization problem~\eqref{eq:Toeplitzproj} using YALMIP~\cite{Lofberg2004YALMIP} in MATLAB, and solved using the MOSEK~\cite{Mosek2019MOSEK} solver. We noticed that the solutions were not accurate when the covariance matrices had very small entries, i.e., on the order of $10^{-11}$. Thus, we normalized the covariance matrices to have unit Frobenius norm before solving~\eqref{eq:Toeplitzproj}. 

\begin{figure}[h!]
\centering
\includegraphics[width=0.45\textwidth]{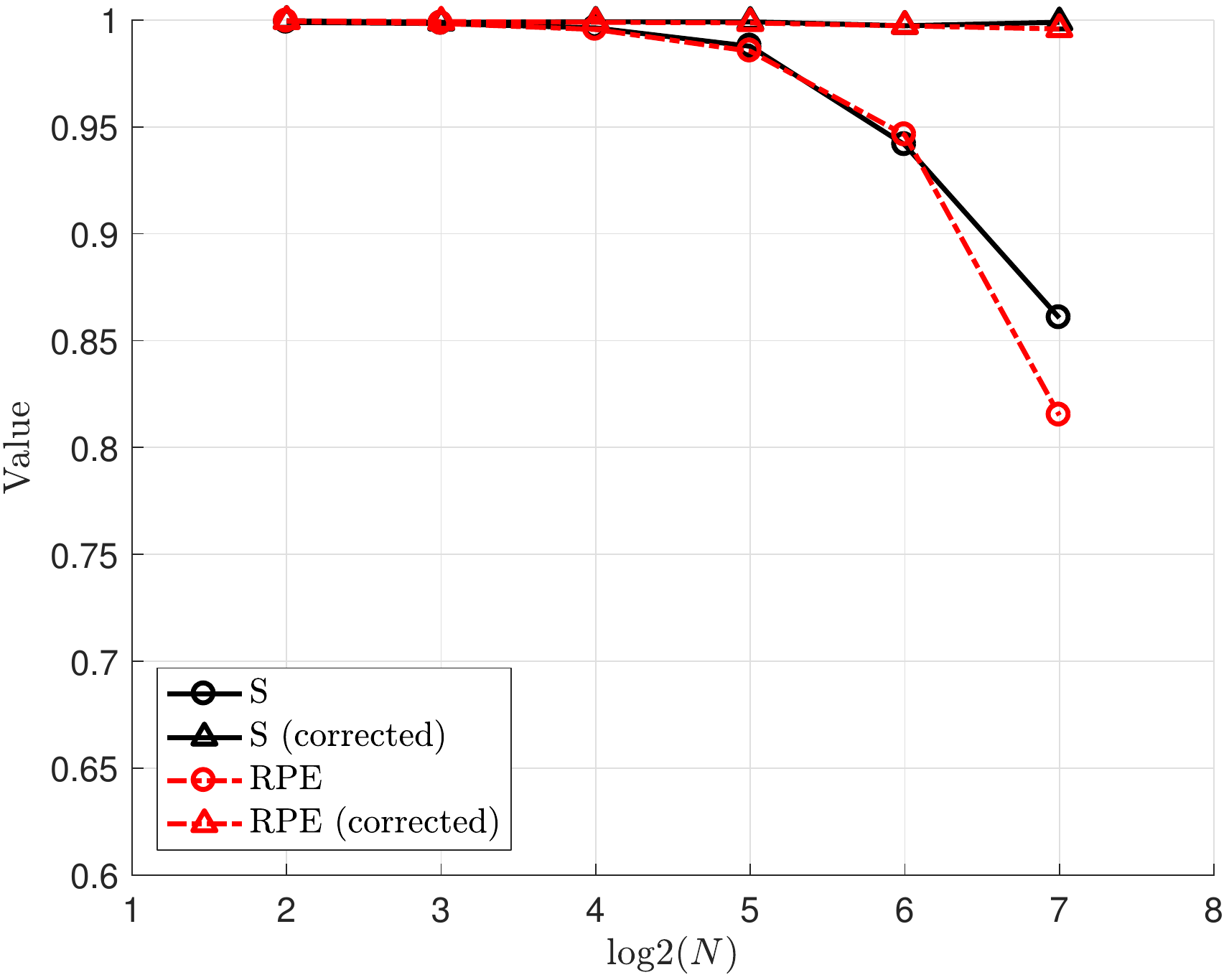}
\caption{The similarity metric $\rmS$ and \ac{RPE} as a function of the number of antennas $N$ for $\Ns=1$. The results are for a simplified case where the radar and communication arrays are collocated at the \ac{RSU} and the vehicle and both systems operate on the same frequency. }
\label{fig:Simmetressingant1}
\end{figure}

\begin{figure}[h!]
\centering
\includegraphics[width=0.45\textwidth]{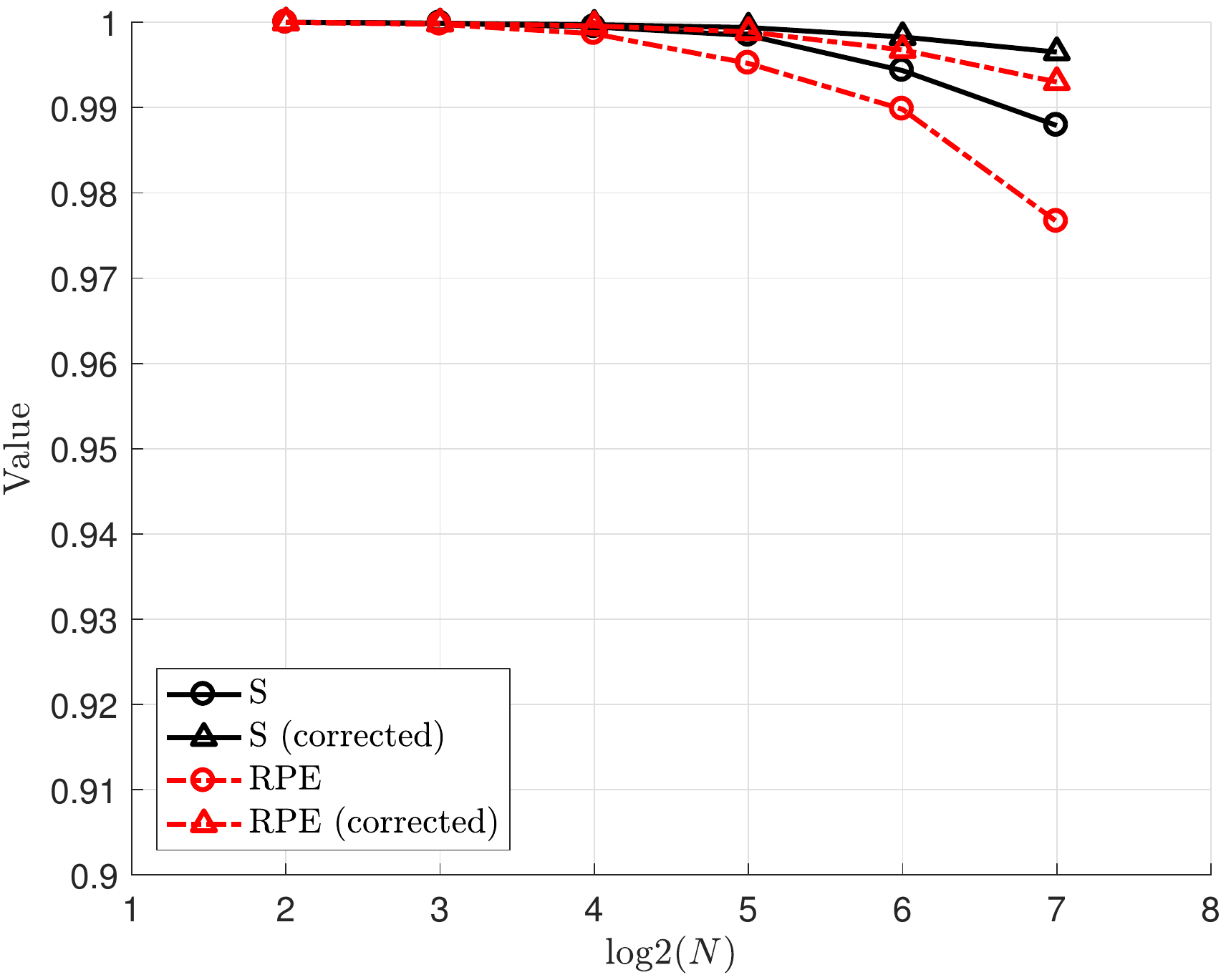}
\caption{The similarity metric $\rmS$ and \ac{RPE} as a function of the number of antennas $N$ for $\Ns=4$. The results are for a simplified case where the radar and communication arrays are collocated at the \ac{RSU} and the vehicle and both systems operate on the same frequency. }
\label{fig:Simmetressingant4}
\end{figure}

We first demonstrate the impact of the bias (discussed in Section~\ref{sec:biascorr}) on the performance of \ac{FMCW} radar and the benefit of bias correction. Note that, in the setup described above, there are several sources of dissimilarity between radar and communication, i.e., different operating frequencies, different locations of the radar and communication antennas on the vehicle, the bias in \ac{FMCW} radar, and the thermal noise. To isolate only the impact of bias, we start with a simple scenario. First, the radar and the communication arrays on the \ac{RSU} are collocated (i.e., horizontally and vertically aligned). We also assume a single antenna array on the vehicle. The radar on the vehicle is colocated with the communication array. This assumption takes away the dissimilarity due to different locations of the antennas. Second, we assume that the radar and communication systems operate in the same band to take out the differences due to the operating frequency. Third, for the first two experiments, we ignore the thermal noise. With this, the only remaining sources of dissimilarity is the bias in \ac{FMCW} radar. We show the similarity in the \ac{APS} of radar and communication as a function of the number of antennas $N$ for ($\Ns=1$) in Fig.~\ref{fig:Simmetressingant1}. We also show the similarity after correcting the bias. In addition, we show the results for \ac{RPE} before and after correction. We can see from the similarity metric results, as well as the \ac{RPE} results, that correcting the bias is helpful as it increases the similarity as well as \ac{RPE}. We show the same result for $\Ns=4$ in Fig.~\ref{fig:Simmetressingant4}. We note that the similarity and \ac{RPE} increase with $\Ns$ as expected (see Section~\ref{sec:congruence} for details). From this result we can also see that bias correction is useful. Therefore, here onwards all the radar results are presented for the corrected case.

\begin{figure}[h!]
\centering
\includegraphics[width=0.45\textwidth]{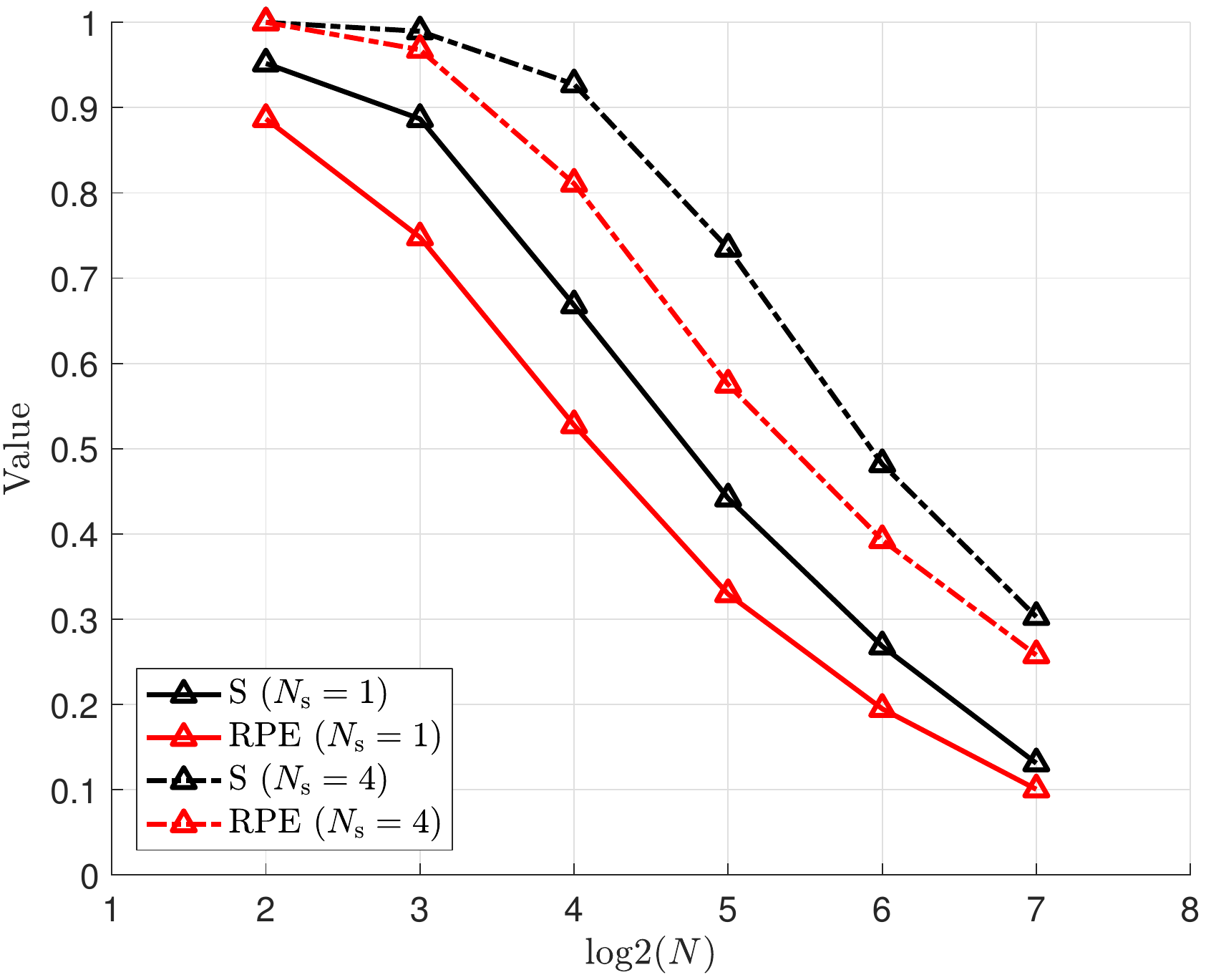}
\caption{The similarity metric $\rmS$ and \ac{RPE} as a function of the number of antennas $N$ for $\Ns=1$ and $\Ns=4$. The results show that the similarity decreases with $N$ and increases with $\Ns$, and further similarity and \ac{RPE} follow the same trend.}
\label{fig:Simmetresallant}
\end{figure}

In the next experiment, we study the similarity of radar and communication in a realistic setup as discussed in the earlier parts of this section. This is to say that now radar and communication have different operating frequencies, different locations of the radar and communication antennas on the vehicle, and there is thermal noise. In this experiment, we study $\Ns=1$ as well as $\Ns=4$ case. As there are four communication arrays on the vehicle, there are four communication channels.
We show the similarity results for the communication channel that has a path with the highest power (typically a \ac{LOS} antenna array). The results of this experiment are shown in Fig.~\ref{fig:Simmetresallant}. We can see that the similarity (and the \ac{RPE}) between communication and radar decreases with the number of antennas $N$. Furthermore, the similarity and \ac{RPE} increase with the number of streams $\Ns$ (this behaviour is expected as discussed in Section~\ref{sec:congruence}). Finally, the similarity metric and \ac{RPE} follow the same trend, i.e., the similarity metric and \ac{RPE} are closely related even for the general off-grid case. 

Now, we conduct an experiment to study the benefit of radar information in establishing an mmWave communication link by using the achievable rate as a performance metric. For this experiment, we consider a single RF-chain at the \ac{RSU}, and one chain per receive array at the vehicle. Therefore only single stream transmission is possible, i.e., $\Ns=1$. We consider that $2$-bit phase-shifters are used at the \ac{RSU} and the vehicle. The radar and the communication system have $128$ antennas at the \ac{RSU}, and there are $16$ antennas in each vehicle array. We use an approximation of \ac{DFT} codebooks based on $2$-bit phase-shifters~\cite{Ali2018Millimeter} at the vehicle and the \ac{RSU}. The $n$th codeword in the RSU codebook is thus a $2$-bit approximation of $\frac{1}{\sqrt{\NRSU}}\baRSU(\arcsin(\frac{2n-\NRSU-1}{\NRSU}))~,n=1,\cdots, \NRSU$. The codebooks for the vehicle are defined in a similar manner. For this experiment, we use rate as a metric. To define the rate, let us say $T_{\rm{train}}$ \ac{OFDM} blocks are used for training, whereas the coherence time of the channel is $T_{\rm{coh.}}$ blocks. Then $(1-\frac{T_{\rm{train}}}{T_{\rm{coh.}}})$ is the fraction of blocks left for data transmission. With this, the rate is
\begin{align}
R=\frac{B_\rmc}{K}\left(1-\frac{T_{\rm{train}}}{T_{\rm{coh.}}}\right) \sum_{k=1}^{K}\log_2 \left(1+\frac{P}{\sigma_\rme^2 K}\sum_{a=1}^A |\bw^{(a)\ast} \bsfH^{(a)}[k] \bff|^2 \right).
\end{align}

For this experiment, we compare three strategies. First, we consider exhaustive-search, in which all pairs of the codewords in the \ac{DFT} codebooks of the \ac{RSU} and the vehicle arrays are evaluated. If we assume that measurements on all the vehicle arrays are made simultaneously, the overhead of exhaustive-search is $\NRSU\times\NV$ \ac{OFDM} blocks. There will be only one beam at the \ac{RSU} that will be used to communicate to all the arrays of the vehicle. As such, at the \ac{RSU}, we select a codeword that provides the highest \ac{SNR} (averaged across all the vehicle arrays). On the vehicle, we choose a codeword for each array that provides the highest \ac{SNR} for the selected codeword at the \ac{RSU}. 

The second strategy is location-assisted. In this strategy, we assume that the location of the vehicle is obtained through the \ac{GNSS}. There is an error of up to $\SI{10}{\meter}$ in the estimated vehicle location~\cite{Garcia2016Location}. Specifically, for each channel realization, we pick an error value uniformly from a disc of $\SI{10}{\meter}$ radius. This vehicle location is communicated to the \ac{RSU} using a low-rate link, e.g., at sub-6 GHz frequencies. Based on the reported location of the vehicle, and the expected error in vehicle location, only a subset of the beams are tried at the \ac{RSU}. Note that, location of the center of the vehicle is reported. As the vehicle has length $\SI{5}{\meter}$, the front and rear antennas are $\pm \SI{2.5}{\meter}$ away from the center. We consider this additional $\SI{2.5}{\meter}$ offset while constructing the subset based on the location information. Specifically, assume that the true angular location of the vehicle - measured from the broadside of the \ac{RSU} array- is $\phi$. Furthermore, assume that the vehicle location estimate available to the \ac{RSU} is $\hat\phi$, and $|\phi-\hat\phi |<\Delta\phi$, where $\Delta\phi$ represents the error of the vehicle localization mechanism. In our case, this error is around $\SI{12.5}{\meter}$. The \ac{RSU} reduces the codebook based on the angular information i.e., $\hat\phi $ and $\Delta\phi$. To formalize this, let us construct an index set $\cL$ such that $n \in \cL$ if  
\begin{align}
\sin(\hat\phi-\Delta\phi)-\frac{1}{\NRSU} \leq \frac{2n-\NRSU-1}{\NRSU},
\label{eq:reduction1}
\end{align}
and 
\begin{align}
\sin(\hat\phi+\Delta\phi)+\frac{1}{\NRSU}\geq  \frac{2n-\NRSU-1}{\NRSU}.
\label{eq:reduction2}
\end{align}
The addition (and subtraction) of $\frac{1}{\NRSU}$ in \eqref{eq:reduction1}~(and~\eqref{eq:reduction2}) ensures that the index set $\cL$ has at least one entry even when $\Delta\phi=0$, i.e., perfect vehicle localization. The above inequalities can be written simply as a compound inequality
\begin{align}
\sin(\hat\phi-\Delta\phi)+1 \leq \frac{2n}{\NRSU} \leq \sin(\hat\phi+\Delta\phi)+1+2/\NRSU.
\label{eq:reduction}
\end{align}

The \ac{RSU} thus only uses the codewords indexed by $\cL$. 

The third strategy is radar-assisted. In this strategy, first, we find the peak in the radar \ac{APS}. Then, we train using a few codewords that point in the directions around the radar \ac{APS} peak. For the rate results, the number of codewords tried for the radar-assisted strategy is the independent variable. 

\begin{figure}[h!]
\centering
\includegraphics[width=0.45\textwidth]{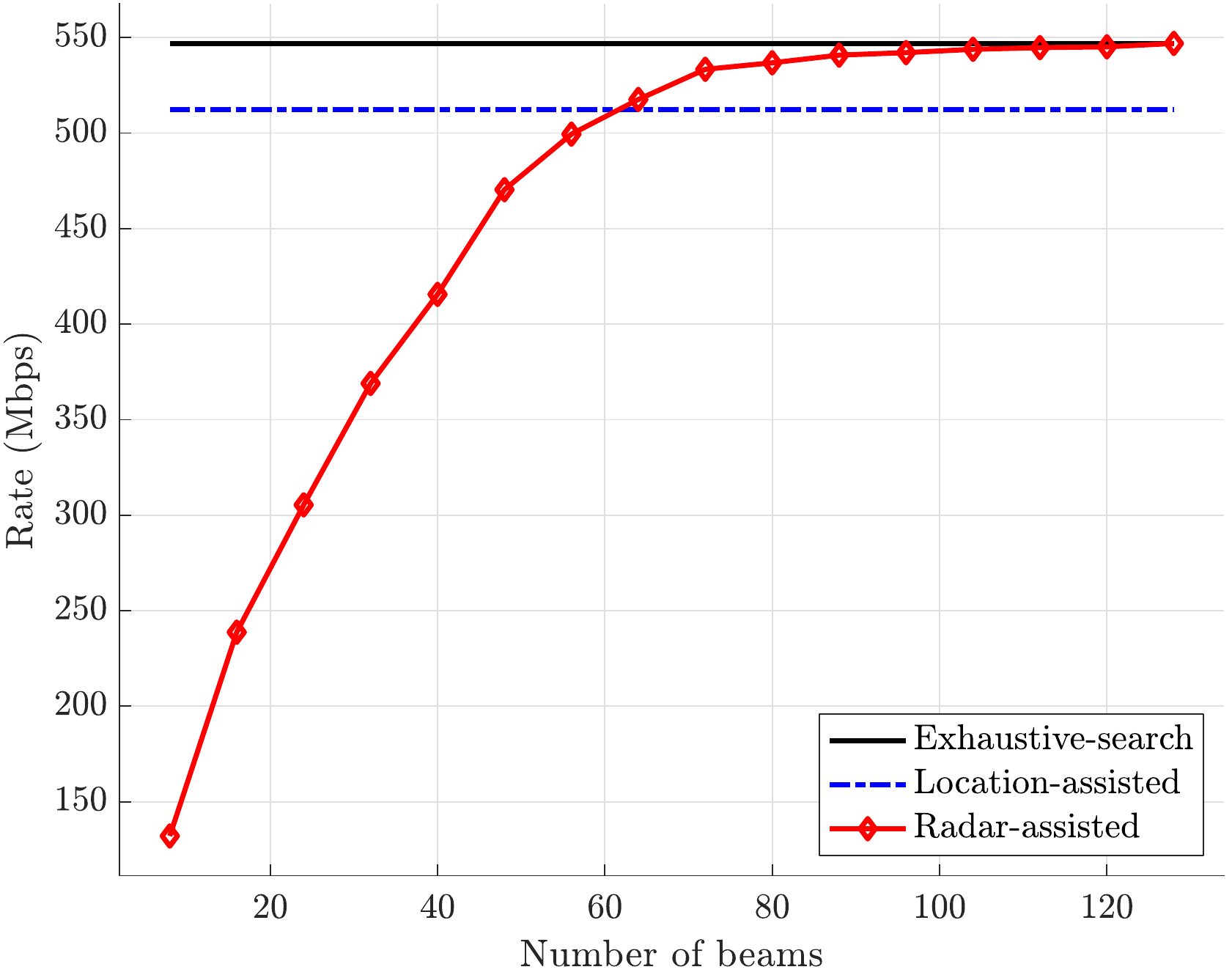}
\caption{Rate versus the number of beams for $T_{\rm{coh.}}\rightarrow\infty$. The proposed radar-assisted strategy achieves the rate of exhaustive-search with fewer measurements, whereas the location-assisted strategy fails to reach the rate of exhaustive search.}
\label{fig:Total_res_Inf}
\end{figure}

\begin{figure}[h!]
\centering
\includegraphics[width=0.45\textwidth]{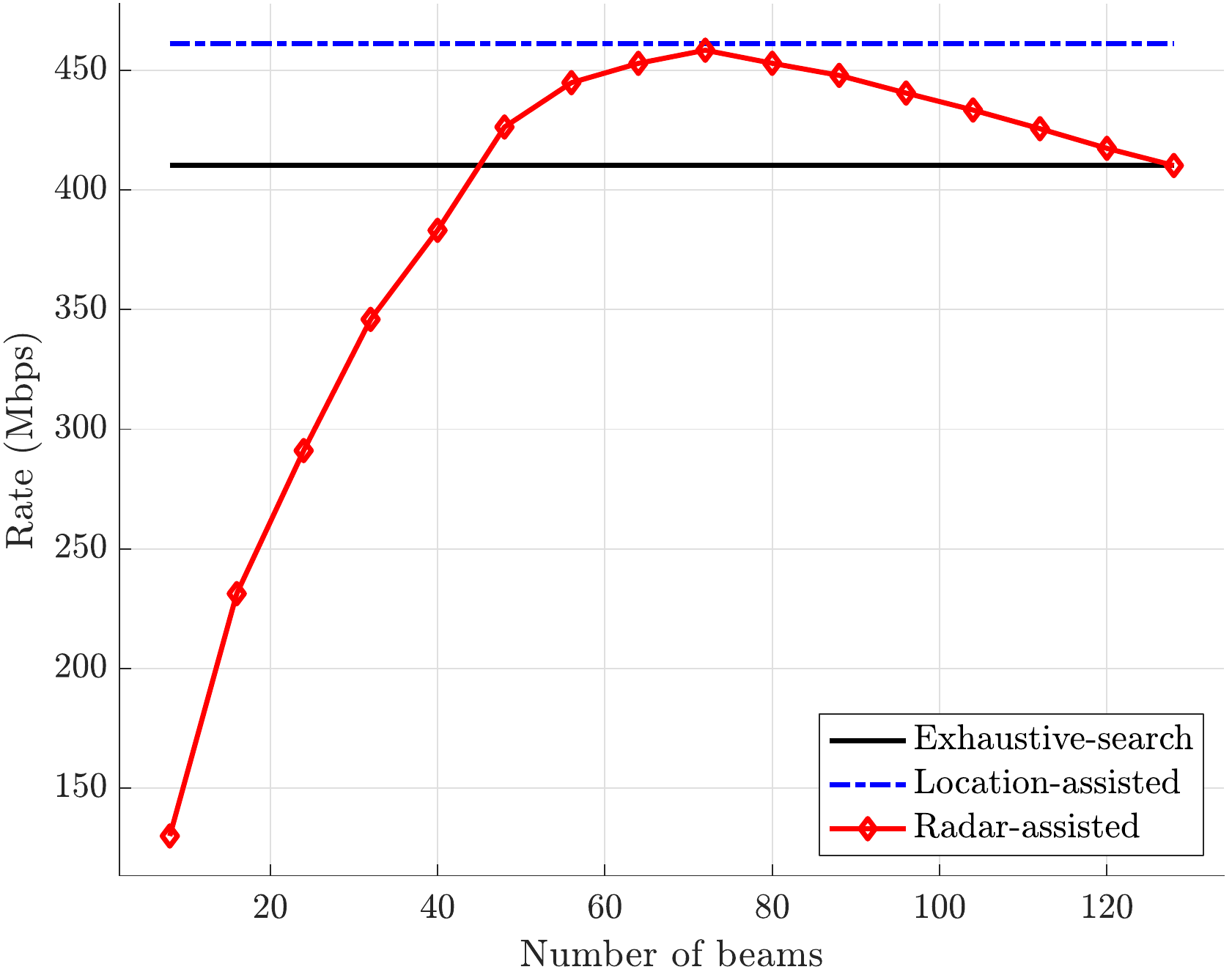}
\caption{Rate versus the number of beams for $T_{\rm{coh.}}=4\NRSU\NV$. The low training overhead radar-assisted and location-assisted strategies have a higher rate than the exhaustive-search strategy.}
\label{fig:Total_res}
\end{figure}

First, we present the results for the case when $T_{\rm{coh.}}\rightarrow\infty$ in Fig.~\ref{fig:Total_res_Inf}. We can see that the radar-assisted strategy can achieve the same rate as exhaustive-search. This rate, however, is achieved by training through $80$ codewords, i.e., around $38\%$ savings in overhead. Note that, on average, location assisted strategy required $51$ codewords, but did not achieve the rate of exhaustive-search. Second, we present the results for a highly dynamic channel with $T_{\rm{coh.}}=4\NRSU\NV$ \ac{OFDM} blocks in Fig.~\ref{fig:Total_res}. Note that, we expect the rate of all the strategies to drop in highly dynamic channels. Strategies with low-overhead, however, are expected to be advantageous in a highly dynamic channel as low training overhead implies a larger duration for the data transmission. The results confirm this observation, as both the location-assisted and radar-assisted strategies perform better than the exhaustive-search. The radar-assisted strategy obtains a rate higher than exhaustive-search with only $50$ measurements, implying an overhead reduction of $60\%$. Note that the rate of the radar-assisted strategy starts to decrease as we keep on increasing the number of measurements. The reason is that once we do enough measurements to find the best beam, additional measurements only increase the overhead and do not improve beam-training. One observation, however, is that the radar-assisted strategy only reaches the rate of the location-assisted strategy. The reason is that the ego-vehicle is in \ac{LOS} with the \ac{RSU} in most of the random drops. In \ac{LOS} channels beam-training based on location information is expected to perform well. 

\begin{figure}[h!]
\centering
\includegraphics[width=0.45\textwidth]{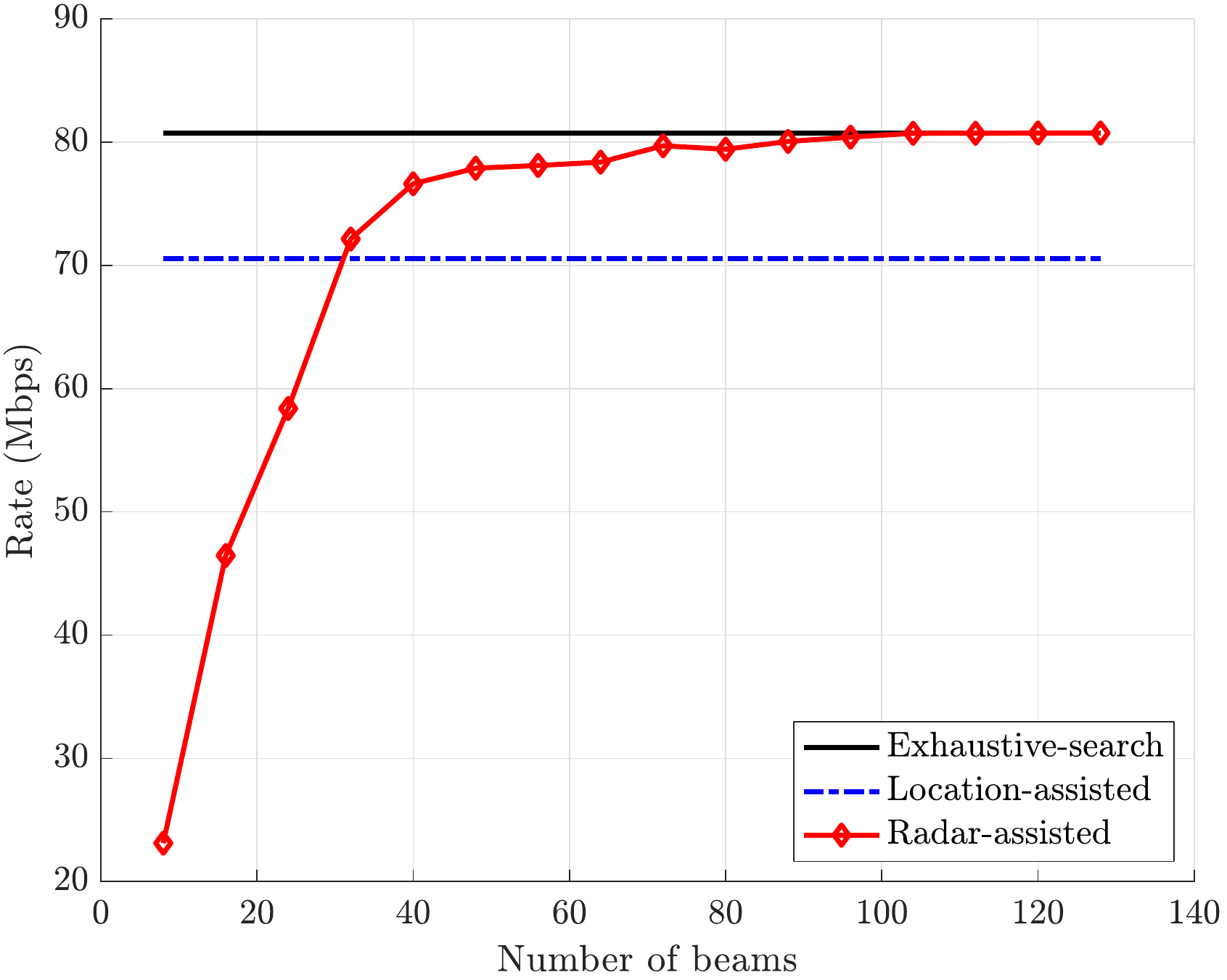}
\caption{Rate versus the number of beams for $T_{\rm{coh.}}\rightarrow\infty$ in \ac{NLOS} channel. The proposed radar-assisted strategy achieves the rate of exhaustive-search with fewer measurements, whereas the location-assisted strategy fails to reach the rate of exhaustive search.}
\label{fig:NLOS_res_Inf}
\end{figure}

\begin{figure}[h!]
\centering
\includegraphics[width=0.45\textwidth]{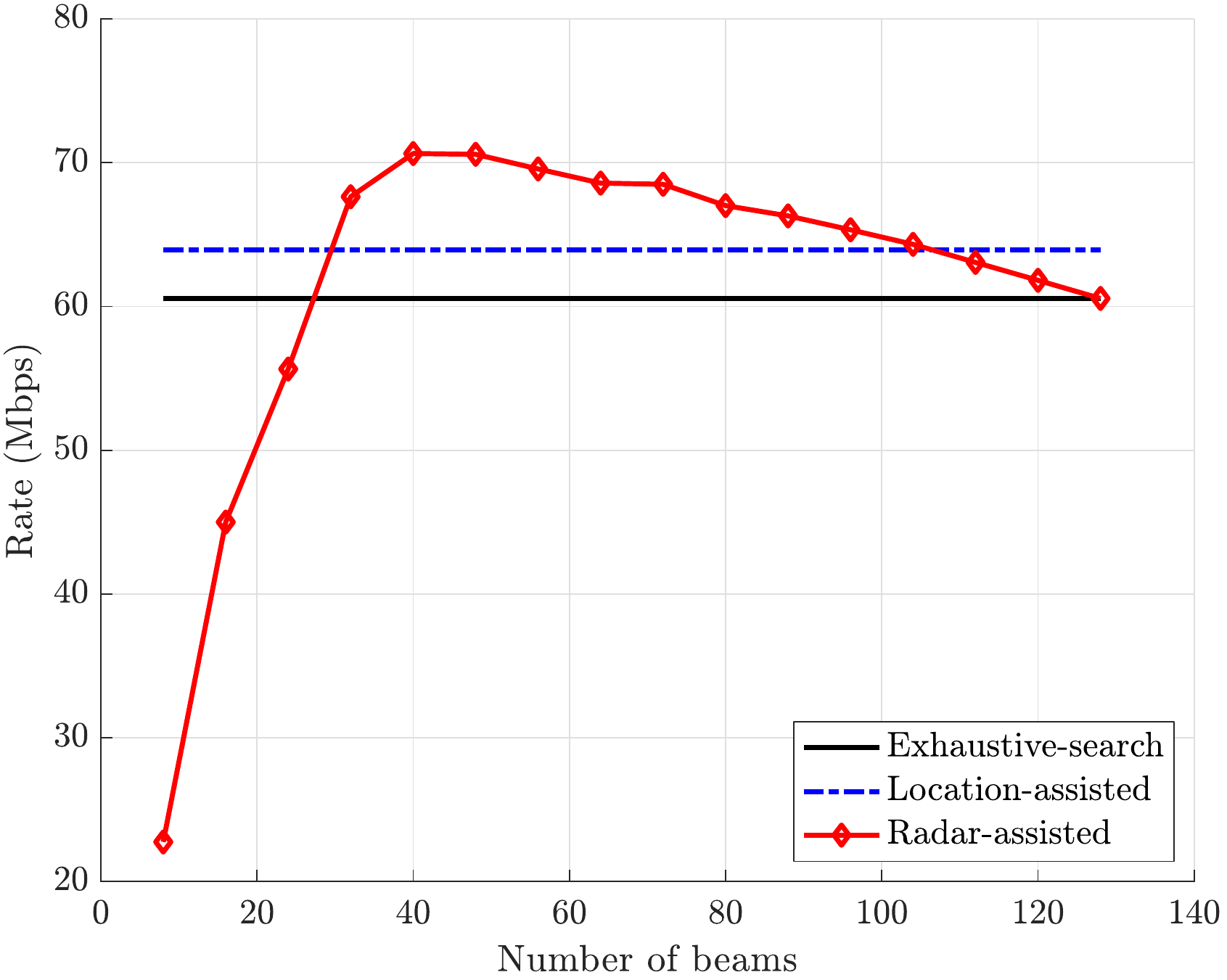}
\caption{Rate versus the number of beams for $T_{\rm{coh.}}=4\NRSU\NV$. The radar-assisted strategy achieves a rate higher than the location-assisted strategy and exhaustive-search with only $30$ measurements, implying a $77\%$ overhead reduction.}
\label{fig:NLOS_res}
\end{figure}

We now study the performance of the proposed strategy in \ac{NLOS} scenario. Specifically speaking, only $179$ out of the $1000$ drops were such that none of the communication antennas on the vehicle had a direct path to the \ac{RSU}. We now present the rate results averaged over these $179$ drops. The results for $T_{\rm{coh.}}\rightarrow\infty$ are presented in Fig.~\ref{fig:NLOS_res_Inf}. We observe that in comparison with the earlier case (i.e., when most of the channels were \ac{LOS}), the radar-assisted strategy can achieve a higher rate than the location-assisted strategy with fewer measurements. Further, the location-assisted strategy fails to achieve the exhaustive-search rate, whereas the radar-assisted strategy achieves almost the exhaustive-search rate with $70$ measurements. The results for $T_{\rm{coh.}}=4\NRSU\NV$  are presented in Fig.~\ref{fig:NLOS_res}. We observe that the radar-assisted strategy can achieve a rate higher than the exhaustive-search and location assisted strategy with only $30$ measurements, implying an overhead reduction of around $77\%$. 

\section{Conclusion and future work}\label{sec:conc}
We used the spatial covariance of the passive radar at the \ac{RSU} to help establish the mmWave communication link. We proposed a simplified radar receiver that does not require the transmit waveform as a reference. Using the proposed architecture, the spatial covariance can be recovered perfectly. Further, we noticed a similarity in the bias that appears in \ac{FMCW} radars to the well-studied problem in \ac{FDD} systems and subsequently used one covariance correction strategy from \ac{FDD} literature to correct the bias in \ac{FMCW} radars. In addition, to compare the similarity of two \ac{APS}, we proposed a similarity metric that is shown to be identical to the \ac{RPE} for the on-grid case, and relates directly to the relative rate. The simulation results based on ray-tracing data showed that bias correction is important and increases the similarity in radar and communication \ac{APS}. The rate results showed that the radar-assisted strategy can reduce the training overhead by around $30-75 \%$ depending on the scenario. Higher gains for the radar-assisted strategy were observed in highly-dynamic channels and in \ac{NLOS} scenarios.

Note that the results for the radar-assisted strategy are based on a simple framework where the potential beam-pairs were pruned based on the radar \ac{APS}. It is, however, possible to use radar covariance in alternative frameworks. For example, spatial covariance of the mmWave communication channel could be estimated using weighted compressed sensing~\cite{Ali2018Spatial}, with the weights coming from radar-covariance. Side information from radar covariance can be incorporated in estimation strategies based on the \ac{AMP} algorithm~\cite{Ma2019approximate}. Finally, we considered the case of a single vehicle with radar transmitters. The impact of multiple vehicles with active radars on the performance of the proposed approach should also be studied.
\bibliographystyle{IEEEtran}
\bibliography{\centrallocation/Abbr,\centrallocation/Master_Bibliography}{}
\end{document}